\documentclass{llncs}
\pdfoutput=1

\usepackage{makeidx}  

\usepackage[utf8]{inputenc}

\usepackage{amsmath}
\usepackage{softdev}
\usepackage[hidelinks]{hyperref}
\usepackage{listings}
\usepackage{booktabs}
\usepackage{multicol}
\usepackage{multirow}
\usepackage{xspace}
\usepackage{multicol}
\usepackage{mdwlist}
\usepackage{graphicx}
\usepackage{caption}
\usepackage{subcaption}

\newcommand{\kalibera}[0]{Kalibera/Jones\xspace}
\newcommand{\unipycation}[0]{Unipycation\xspace}
\newcommand{\swithon}[0]{CPython-SWI\xspace}
\newcommand{\swipy}[0]{PyPy-SWI\xspace}
\newcommand{\jythog}[0]{Jython-tuProlog\xspace}
\newcommand{\pyprolog}[0]{\emph{Python} $\rightarrow$ \emph{Prolog}\xspace}
\newcommand{\pyprolognc}[0]{\emph{Python} $\overset{nc}{\rightarrow}$ \emph{Prolog}\xspace}

\begin{document}

\frontmatter
\pagestyle{headings}  

\mainmatter              

\title{Approaches to Interpreter Composition}
\titlerunning{Approaches to Interpreter Composition}  
\author{%
	Edd Barrett \and
	Carl Friedrich Bolz \and
	Laurence Tratt
}

\newcommand{\kcl}{Software Development Team, Informatics, King's College London, UK}
\institute{
	\url{http://eddbarrett.co.uk/}\qquad%
	\url{http://cfbolz.de/}\qquad%
	\url{http://tratt.net/laurie/}\\%
	\vspace{1em}\kcl%
}

\maketitle

\begin{abstract}
In this paper, we compose six different Python and Prolog VMs into 4
pairwise compositions: one using C interpreters; one running on the JVM; one
using meta-tracing interpreters; and one using a C interpreter and a
meta-tracing interpreter. We show that programs that cross the language
barrier frequently execute faster in a meta-tracing composition, and that
meta-tracing imposes a significantly lower overhead on composed programs
relative to mono-language programs.

\end{abstract}

\bibliographystyle{splncs03}







\section{Overview}

Programming language composition aims to allow the mixing of programming languages in a
fine-grained manner. This vision brings many challenging problems, from
the interaction of language semantics to performance. In
this paper, we investigate the runtime performance of composed
programs in high-level languages.
We start from the assumption that execution of such programs is most likely to
be through composing language implementations that use interpreters and VMs,
rather than traditional compilers. This raises the question: how do different styles
of composition affect performance? Clearly, such a question cannot have a
single answer but, to the best of our knowledge, this issue has not been
explored in the past.

This paper's hypothesis is that meta-tracing -- a relatively new technique used
to produce JIT (Just-In-Time) compilers from interpreters~\cite{bolz14impact} --
will lead to faster interpreter composition than traditional approaches.
To test this hypothesis, we present a Python and Prolog composition which allows
Python programs to embed and call Prolog programs. We then implement the
composition in four different ways, comparing the absolute times and the
relative cross-language costs of each. In addition to the `traditional'
approaches to composing interpreters (in C and upon the JVM), we also
investigate the application of meta-tracing to interpreter composition.
The experiments we then carry out confirm our initial hypothesis.

There is a long tradition of composing Prolog with other languages (with
e.g.~Icon~\cite{lapaime86logicon}, Lisp~\cite{robinson80loglisp}, and
Smalltalk~\cite{gybels03soul}) because one can express certain types of programs
far easier in Prolog than in other languages. We
have two additional reasons for choosing this pairing. First, these languages represent
very different points in the language design space: Prolog's inherently
backtracking nature, for example, leads to a much different
execution model than that of Python.
It is therefore reasonable to assume that it is harder to optimise
the composition of such languages than it would be if they were substantially
similar. Second, each language has incarnations in several
different language implementation paradigms, making our experiment practical. Few
other languages meet both these criteria at the current time.

The four compositions we implemented are as follows: \emph{\swithon} composes two C
interpreters; \emph{\swipy} composes RPython and C interpreters;
\emph{\unipycation} composes two RPython interpreters; and \emph{\jythog} two JVM
interpreters. \swithon, \swipy, and \jythog represent `traditional' techniques
to language composition; \unipycation embodies a new technique. Each composition implements exactly the
same user-visible language, so composed programs run without change on each
of the composed VMs. We have implemented the Python to Prolog interface as
a mixture of pure Python code and code written in the language of the underlying
system (RPython, C, Java), deliberately modelling how such libraries are typically
implemented. \unipycation is an extension of work presented in a previous workshop
paper~\cite{barrett13unipycation}, which we have since enhanced
to allow full cross-language inlining. Note however, that unlike our previous work, this paper
restricts itself to Python code calling Prolog, and not vice versa (some of
the composed VMs are fundamentally unable to support the latter).

To address our hypothesis, we present a series of microbenchmarks, each designed
to focus on a single aspect of cross-language performance, as well as a small
set of (relatively) larger benchmarks which cross language barriers more freely.
Each benchmark has both mono-language variants written in either Python or
Prolog, and composed variants written in a mixture of Python and Prolog. We
report our benchmark timings in absolute seconds, but also as ratios of
the mono-language variants and \unipycation. The latter results allow us to understand the cost of
moving from mono-language to composed variants, as well as the performance consequences
of using meta-tracing for composed VMs. To ensure that our results are as
robust as possible, we use the rigorous methodology of \kalibera; since
we are, as far as we know, the first `users' of this methodology, we also
describe our experiences of it.

Our results break down as follows. First, as an unsurprising base-line,
benchmarks which rarely cross the language barrier are dominated by the performance
characteristics of the individual language implementations; put another way, the
particular composition technique used makes little difference. Second, for
those benchmarks which cross the language barrier frequently, \unipycation is,
on average, significantly faster than \swipy; \swithon is somewhat slower again;
and \jythog is hugely (unusably) slower. Third, \unipycation executes composed
microbenchmarks in the same performance ballpark as the mono-language variants
(at worst, well within an order of magnitude difference; and often only a few
percent slower). By contrast, the next best performing composed VM \swipy is
often at least 1--3 orders of magnitude worse than the mono-language variants.
Bearing in mind the dangers of generalising from any specific set of benchmarks,
we consider these results to validate the paper's hypothesis.

Our contributions are as follows:
\begin{itemize*}
\item We present a viable Python/Prolog composition and show four different implementations each using a different composition style.
\item We present the first experiment designed to help understand the effects
    of different composition styles upon performance.
\item We thoroughly analyse and discuss the results of the experiment, breaking down 
    the impact that each composition style has on performance.
\end{itemize*}

\noindent The experiments contained in this paper are downloadable and repeatable from:%
\begin{center}
    \url{http://soft-dev.org/pubs/files/interpcomp/}\label{download}
\end{center}

\noindent This paper is structured as follows. We briefly introduce meta-tracing
(Section~\ref{sec:metatrace}) and Prolog for those readers who may be
unfamiliar, or rusty, with either. After introducing the generic \unipycation
interface, including an example Python and Prolog composition with an
implementation of Connect 4 (Section~\ref{sec:demonstration}), we then describe
the four implementations of this interface (Section~\ref{sec:vms}). The
experimental setup (Section~\ref{sec:experiment}) is followed by a quantitative
analysis of its results (Section~\ref{sec:results}) and a qualitative
discussion (Section~\ref{sec:discussion}).

\section{Meta-tracing}
\label{sec:metatrace}

This section briefly introduces the concept of meta-tracing.  Meta-tracing takes
as input an interpreter, and from it creates a VM containing both the interpreter and a tracing JIT
compiler~\cite{mitchell70design,sullivan03dynamic,bolz09tracing,yermolovich09optimization,bebenita10spur}.
Although tracing is not a new idea (see~\cite{bala00dynamo,gal06hotpathvm}), it
traditionally required manually implementing both interpreter and trace
compiler. Meta-tracing, in contrast, automatically generates a trace compiler
from an interpreter.
At run-time, user programs running in the VM begin their execution in the
interpreter.  When a `hot loop' in the user program is encountered, the actions
of the interpreter are traced (i.e.~recorded), optimised, and converted to
machine code. Subsequent executions of the loop then use the fast machine code
version rather than the slow interpreter. Guards are left behind in the machine
code so that execution can revert back to the interpreter when the path recorded
by the trace differs from that required.

Meta-tracing works because of the particular nature of interpreters.
Irrespective of whether they operate on bytecode or ASTs, are iterative or recursive,
interpreters are fundamentally a large loop:
`load the next instruction; perform the associated actions; repeat the loop'. To
generate a tracing JIT, the language implementer annotates the interpreter to
inform the meta-tracing system when a loop\footnote{Loops are often, though not
exclusively, program counter jumps with a negative index.} at position $pc$
(program counter) has been encountered; the meta-tracing system then decides if
the loop has been encountered often enough to start tracing. The annotation also
tells the meta-tracing system that execution of the program at position $pc$ is
about to begin and that if a machine code version is available, it should be
used; if not, the standard interpreter will be used.

The main extant meta-tracing language is RPython, a statically-typed subset of
Python which translates to C. RPython's type system is similar to Java's,
extended with further analysis, e.g.~to ensure that list indices are not
negative. Users can influence the analysis with \texttt{assert} statements, but
otherwise it is fully automatic. Unlike seemingly similar languages
(e.g.~Slang~\cite{ingalls97back} or PreScheme \cite{kelsey94tractable}), RPython
is more than just a thin layer over C: it is, for example, fully garbage
collected and has several high-level datatypes (e.g.~lists and dictionaries).
Despite this, VMs written in RPython have performance levels which far exceed traditional
interpreter-only implementations~\cite{bolz14impact}.

Meta-tracing is appealing for composed VMs because it holds the prospect of
achieving meaningful cross-language optimisations for little or no effort. We
explore this in the context of \unipycation in Section~\ref{sec:unipycation}.

\section{Prolog Background}
\label{sec:prolog}

While we assume that most readers have a working knowledge of a mainstream
imperative language such that they can understand Python, we can not reasonably make the same
assumption about Prolog.  This section serves as a brief introduction to
Prolog for unfamiliar readers (see e.g.~\cite{bratko01prolog} for more
details). Those familiar with Prolog will notice a distinct imperative
flavour to our explanations. This is intentional, given the paper's likely
audience, but nothing we write should be considered as precluding the
logic-based view of Prolog.

Prolog is a rule-based logic programming language whose programs consist of a
database of predicates which is then queried. A predicate is related to, but subtly
different from, the traditional programming language concept of a function.
Predicates can be loosely thought of as overloaded pattern-matching functions
that can generate a stream of solutions (including no solutions at all). Given
a database, a user can then query it to ascertain the truth of an expression.

Prolog supports the following data types:

\begin{description*}
\item[Numeric constants] Integers and floats.
\item[Atoms] Identifiers starting with a lowercase letter e.g.~\texttt{chair}.
\item[Terms] Composite structures beginning with a lowercase letter
e.g.~\texttt{vector(1.4, 9.0)}. The name (e.g.~\texttt{vector}) is the term's
\emph{functor}, the items within parentheses its \emph{arguments}.
\item[Variables] Identifiers beginning with either an uppercase letter
(e.g.~\texttt{Person}) or an underscore. A variable denotes an as yet unknown
value that may become known later when a concrete value is \emph{bound} to
the variable.
\item[Lists] Lists are made out of cons cells, which are simply terms of the
functor \texttt{'.'}. Since lists are common, and the
\texttt{'.'} syntax rather verbose, lists can be expressed using a comma
separated sequence of elements enclosed inside square brackets. For example,
the list \texttt{[1,2,3]} is equivalent to \texttt{'.'(1, '.'(2, '.'(3,
[])))}. Furthermore, a list can be denoted in terms of its \emph{head} and
\emph{tail}, for example \texttt{[1 | [2, 3]]} is equivalent to
\texttt{[1,2,3]}.
\end{description*}

\noindent To demonstrate some of these concepts, consider the Prolog rule database shown in
Listing~\ref{lst:path}. The \texttt{edge} predicate describes a directed
graph which may, for example, represent a transit system such as the London
Underground. The \texttt{path} predicate accepts four arguments
and describes
valid paths of length \texttt{MaxLen} or under, from the node \texttt{From} to
the node \texttt{To}, as the list \texttt{Nodes}. In this example,
\texttt{a}, \texttt{b}, \ldots, \texttt{g} are atoms. The
expression \texttt{edge(a, c)} defines a predicate called \texttt{edge} which is true when the atoms \texttt{a} then \texttt{c} are passed as arguments.
The expression \texttt{path(From, To, MaxLen, Nodes,
1)} is a call to the \texttt{path} predicate passing as arguments, four variables and a integer constant.

\begin{lstlisting}[caption={A Prolog rule database.}, label={lst:path}]
edge(a, c). edge(c, b). edge(c, d). edge(d, e).
edge(b, e). edge(c, f). edge(f, g). edge(e, g).
edge(g, b).

path(From, To, MaxLen, Nodes) :-
    path(From, To, MaxLen, Nodes, 1).

path(Node, Node, _MaxLen, [Node], _Len).
path(From, To, MaxLen, [From | Ahead ], Len) :-
    Len < MaxLen, edge(From, Next),
    Len1 is Len + 1,
    path(Next, To, MaxLen, Ahead, Len1).

\end{lstlisting}

Queries can either succeed or fail.  For example, running the query \texttt{edge(c,
b)} (``is it possible to transition from node \texttt{c} to node
\texttt{b}?'') against the above database succeeds, but
\texttt{edge(e, f)} (``is it possible to transition from node
\texttt{e} to
node \texttt{f}?'') fails.  When a
query contains a variable, Prolog searches for solutions,
binding values to variables. For example, the query \texttt{edge(f, Node)} (``which node
can I transition to from \texttt{f}?'') binds \texttt{Node} to the
atom \texttt{g}. Queries can produce multiple solutions. For example,
\texttt{path(a, g, 7, Nodes)} (``Give me
paths from \texttt{a} to \texttt{g} of maximum length 7'') finds
several bindings for \texttt{Nodes}:
\texttt{[a, c, b, e, g]}, \texttt{[a, c, d, e, g]},
\texttt{[a, c, f, g]}, and \texttt{[a, c, f, g, b, e, g]}.

Solutions are enumerated by recording \emph{choice points} where more
than one rule is applicable. If a user requests another solution, or if an
evaluation path fails, Prolog backtracks and explores alternative search
paths by taking different decisions at choice points. In the above example, \texttt{edge(From, Next)} (line 10)
can introduce a choice point, as there can be several ways
of transitioning from one node to the next.

\section{The \unipycation interface}
\label{sec:uniprogs}

\unipycation is both an implementation of a Python and Prolog composition, and
an interface which other implementations can target. In this paper we provide
4 composed VMs which, while they differ significantly internally, presents the
same interface to the programmer. All \unipycation programs start with Python
code, subsequently using the \texttt{uni} module to gain access to the Prolog
interpreter, as in the following \unipycation program:
\begin{lstlisting}[label={lst:basic_api_usage}, caption={Program showing basic usage of the \texttt{uni} API.}]
import uni
e = uni.Engine("""
    % parent(ParentName, ChildName, ChildGender)
    parent(jim, lucy, f). parent(evan, lily, f).
    parent(jim, sara, f).  parent(evan, emma, f).
    parent(evan, kyle, m).

    siblings(Person1, Person2, Gender) :-
        parent(Parent, Person1, Gender),
        parent(Parent, Person2, Gender),
        Person1 \= Person2.
""")

for (p1, p2) in e.db.siblings.iter(None, None, "f"):
    print "%s and %s are sisters" % (p1, p2)
\end{lstlisting}
The Python code first imports the \texttt{uni} module and creates an
instance of a Prolog engine (line 2). In this case, the engine's database is
passed as a Python string (lines 3-11 inclusive), although
databases can be loaded by other means (e.g.~from a file). The Prolog database
models a family tree with a series of \texttt{parent} facts. The
\texttt{siblings} predicate can then be used to infer pairs of same-sex
siblings. This example queries the Prolog database for all pairs of sisters
(line 14) and successively prints them out (line 15). The Prolog database is accessed
through the \texttt{db} attribute of the engine (e.g.~\texttt{e.db.siblings}).
Calling a Prolog predicate as if it were a Python method transfers execution to
Prolog and returns the first solution;
if no solution is found then \texttt{None} is returned. Predicates also have
an \texttt{iter} method which behaves as a Python iterator, successively generating all solutions (as shown in the example on line 14). Solutions returned by the iterator interface are generated in a lazy on-demand fashion.

Calling to, and returning from, Prolog requires datatype conversions.
Python numbers map to their Prolog equivalents; Python strings
map to Prolog atoms; and Python (array-based) lists map to Prolog
(cons cell) lists. A Python \texttt{None} is used to represent a (distinct) unbound
variable. Composite terms (not shown in the example) are passed as
instances of a special Python type \texttt{uni.Term}, instantiated by a
call to a method of the engine's \texttt{terms} attribute. For example,
a term equivalent to \texttt{vec(4, 6)} is instantiated by executing
\texttt{e.terms.vec(4, 6)}.  Prolog solutions are returned to Python as
a tuple, with one element for each \texttt{None} argument passed.

When run, Listing~\ref{lst:basic_api_usage} generates the following output:
\begin{lstlisting}
lucy and sara are sisters
lily and emma are sisters
sara and lucy are sisters
emma and lily are sisters
\end{lstlisting}

\subsection{High-Level and Low-Level Interface}

Although it may not be evident from studying Unipycation code, our composed
VMs are implemented in two distinct levels.  The \texttt{uni} module
(shown in the previous section) exposes a high-level programming interface
written in pure Python. This allows us to use it unchanged
across our four composed VMs. At a lower-level, each of the
compositions exposes a fixed API for the \texttt{uni} module to interface with. This split has
several advantages: it maximises code sharing between the compositions;
makes it easier to experiment with different \texttt{uni} interface
designs; and models a typical approach used by many Python
libraries that wrap non-Python code such as C libraries.

The low-level interface provides a means to create a Prolog engine with a
database. It also provides an iterator-style query interface and some basic
type conversions---higher-level conversions are delegated to the \texttt{uni} module.
In particular, Prolog terms have the thinnest possible wrapper put
around them. When returning solutions to Python, the high-level interface
must perform a deep copy on terms, as most of the Prolog implementations garbage
collect, or modify, them when a query is pumped for more solutions.

The low-level interface is an implementation detail that is not intended
for end-users. Failing to follow certain
restrictions leads to undesirable behaviour, from returning arbitrary
results to segmentation faults (depending on the composed VM). The \texttt{uni}
module follows these restrictions by design.

\subsection{An Example}
\label{sec:demonstration}

To show what sorts of programs one might wish to write with the \unipycation
interface, we present a small case study of a Connect 4 implementation. Connect
4 is a well known strategy game, first released in 1974. The
game involves two players (red and yellow), and a vertically standing
grid-like board, divided into 6 rows and 7 columns.  Players take turns
dropping one of their coloured counters into a column. The first player
to place 4 counters in a contiguous horizontal, vertical, or diagonal
line wins.

Connect 4 has two distinct aspects: a user interface and an AI player.  Without
wishing to start a war between fans of either language, we humbly suggest that
Python is better suited to expressing the GUI, whereas Prolog is better suited
to expressing the behaviour of an AI player. Our implementation follows this
pattern, and has a Python GUI of 183LoC and a Prolog AI player of 168LoC.

\begin{figure}[tb]
\centering
\includegraphics[width=.3\textwidth]{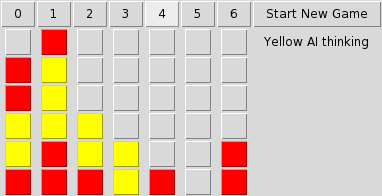}
\caption{The Connect 4 GUI using Python/Tk.}
\end{figure}

The basic operation of the case study is as follows.  The Python part of the program
performs all interactions with users and stores the state of the board.
After every move, the Python part divides the
state of the board into two lists (\texttt{reds} and \texttt{yellows}),
encoding counter positions as Prolog terms (e.g.~a counter at row one, column
two is encoded as \texttt{c(2, 1)}).
Unipycation's Python $\rightarrow$ Prolog interface
is then used to query the Prolog \texttt{has\_won}
predicate passing as arguments the two lists and a final unbound variable. If the Prolog
interpreter finds a binding to the variable, then the game has finished, and the last player to move
has won.

If no player has won, and the next player to move is the AI opponent, the
Python part hands over to the Prolog AI player. To decide a
good move, the computer opponent uses a bounded-depth minimax
solver~\cite{neumann28theorie,neumann44theory} implemented efficiently
using alpha-beta pruning~\cite{edwards61alpha}; we used Bratko's alpha-beta
framework as a reference implementation~\cite[p.~586]{bratko01prolog}.
This approach considers the game as a tree of potential moves, where each
move is characterised by the positions of the counters (again as two lists)
and whose move it is next. Each move has an associated cost which is
used as a basis for deciding a good move. We model Connect 4 as a zero sum
game, so for one player a good move minimises the cost,
whereas for the other player a good move maximises the cost.
The alpha-beta framework requires us to define three
predicates: one to tell the framework whose turn it is in a given position,
one to calculate all possible next moves, and one to
calculate the cost of a move. Given these predicates, the alpha-beta
framework can make an informed decision about the best move for the computer
opponent. Once the best move has been chosen, it is passed back to Python and
the game state is updated to reflect the AI player's move.

We discuss our experiences of using Unipycation for the case study in
Section~\ref{sec:discussion}.

\section{The Composed VMs}
\label{sec:vms}

In this section we describe our four composed VMs. Each takes two existing
language implementations -- one Python, and one Prolog -- and glues them
together. The language implementations and the versions used, are shown in
Table~\ref{table:langimps}; the size of the glue is shown in
Table~\ref{tab:the_vms}. To ensure our experiment is repeatable,
we used only open source language implementations.
\begin{table}[tb]
\begin{center}
    \begin{subtable}{.4\textwidth}
    \begin{tabular}{ll}
        \toprule
        Implementation & Version \\
        \midrule
        CPython        & 2.7.6 \\
        Jython         & HG aa042b69bdda \\
        SWI Prolog     & 6.6.1 \\
        PyPy           & HG fa985109fcb9\\
        Pyrolog        & HG cab0c5802fe9\\
        tuProlog       & SVN 1263\\
        \bottomrule
    \end{tabular}
    \caption{}
    \label{table:langimps}
    \end{subtable}

    \begin{subtable}{.5\textwidth}
    \begin{tabular}{llll}
        \toprule
        VM              & LoC\\
        \midrule
        \swithon        & 800 (Python)\\
        \swipy           & See \swithon (same code)\\
        \unipycation     & 650 (RPython)\\
        \jythog          & 350 (Python)\\
        \bottomrule
    \end{tabular}
    \caption{}
    \label{tab:the_vms}
    \end{subtable}
\end{center}
    \caption{(a) Base language implementations used, and (b) lines of glue code for each of the composed VMs.}
\end{table}

\subsection{\swithon}

\swithon composes two interpreters written in C: CPython and SWI Prolog~\cite{wielemaker12future}.
CPython is the de-facto standard implementation of Python and
SWI Prolog is the most popular open-source Prolog interpreter.

\swithon's glue code is the largest of the composed VMs.
This reflects the intricacies of mixing together
low-level interpreters with different garbage collectors (etc.). To make our
life somewhat easier, the glue code was written in Python using the CFFI
library.\footnote{\url{https://cffi.readthedocs.org/}}
On first import, CFFI emits C code which is
compiled into a CPython extension and loaded; the extension in turn loads
SWI Prolog as a dynamic library, meaning, in essence, that the
composition occurs at the C level. Around 430LoC relate to CFFI from which
around 2KLoC of C code are generated. To give a flavour of our
implementation style, Listing~\ref{lst:swithon_conv} shows an excerpt from
the \swithon conversion code (conversion code for the other VMs looks much
the same).  Implementing \swithon required no modifications to either
CPython or SWI Prolog.

\begin{lstlisting}[caption={Prolog $\rightarrow$ Python conversion code in \swithon},label={lst:swithon_conv}]
def py_of_pro(pro_term):
    lib = C._libpl
    pl_type = lib.PL_term_type(pro_term)
    if pl_type == lib.PL_ATOM:
        return py_str_of_pro_atom(pro_term)
    elif pl_type == lib.PL_TERM:
        return py_term_of_pro_term(pro_term)
    elif pl_type == ...:
        ...

def py_str_of_pro_atom(pro_atom):
    return C.get_str(pro_atom)

def py_term_of_pro_term(pro_term):
    return CoreTerm.from_pl_term(pro_term)
\end{lstlisting}

\subsection{\swipy}

\swipy is a simple variant of \swithon, swapping CPython for PyPy.
Because we used CFFI for \swithon's glue
code, `creating' \swipy is simply a matter of invoking the same (unchanged) code under PyPy.
\swipy is interesting in that it composes a JITing VM
(PyPy) with a traditional interpreter (SWI Prolog).

\subsection{\unipycation}
\label{sec:unipycation}

\unipycation composes two RPython (see Section~\ref{sec:metatrace}) interpreters:
PyPy~\cite{rigo06pypy} and Pyrolog~\cite{bolz10towards}. PyPy is a
fast~\cite{bolz14impact}, industrial strength implementation of Python 2.7.3 and
has been widely adopted in the open-source community. Pyrolog has
its origins in a research project investigating the application of meta-tracing
to logic programming. As PyPy modules need to be statically compiled into the VM, the
low-level interface required adding a module to PyPy's source. Pyrolog is, from
this paper's perspective, unmodified.\footnote{Calling Python from Prolog
required modifying Pyrolog, but we do not tackle that issue in this paper.}
In the rest of this subsection, we explain how meta-tracing enables
cross-language optimisations, and how \unipycation can inline Prolog code called
from Python.

Each of \unipycation's constituent interpreters has its own meta-tracing
JIT, such that running Python or Prolog code under \unipycation is
as fast as running it in a stand-alone PyPy or Pyrolog VM. Most of
\unipycation's
optimisations are inherent to meta-tracing. Both
PyPy and Pyrolog are optimised for meta-tracing in the sense that their
implementation has, where necessary, been structured according to
meta-tracing's demands. Such structuring is relatively minor
(see~\cite{bolz14impact} for more details): most
commonly, tracing annotations are added to the
interpreter to provide hints to the tracer (e.g.~``this RPython function's loops
can safely be unrolled''). Less commonly, code is
e.g.~moved into its own function to allow an annotation to be applied. To improve
the performance of \unipycation, we added ten such annotations, but PyPy and
Pyrolog themselves were mostly left unchanged. This is not to suggest that the
two interpreters have identical execution models: PyPy is a naive recursive
bytecode interpreter, whereas Pyrolog uses continuation-passing style (in fact,
Pyrolog uses two continuations: one for success and one for
failure~\cite{bolz10towards}
using a trampoline).

A tracing JIT's natural tendency to aggressively type-spe\-cial\-ise code
(see~\cite{gal09trace}) is important in reducing the
overhead of object conversions between interpreters. Tracing a call from one
language to the other naturally traces the object conversion code in
the \texttt{uni} module, type specialising it. In essence, \unipycation
assumes that the types of objects converted at a given point in the code
will stay relatively constant; similar assumptions are the basis of all
dynamically typed language JITs. Type specialisation leaves behind a type
guard, so that if the assumption of type constancy is later invalidated, execution returns
to the interpreter. If, as is likely, the object conversion is part of a
bigger trace with a previous equivalent type guard, RPython's tracing optimiser will
remove the redundant type guard from the object conversion.

A related optimisation occurs on the wrapped objects that are created by
cross-interpreter object conversion. The frequent passing of objects between the two
interpreters would seem to be inefficient, as most will create a new
wrapper object each time they are passed to the other interpreter. Most such
objects exist only within a single trace and their allocation costs are
removed by RPython's escape analysis~\cite{bolz11allocation}.

The final optimisation we hope to inherit from meta-tracing is
inlining. While it is relatively easy to inline Python
code called from Prolog, the other way around -- which is of rather more
importance to this paper -- is complicated by tracing pragmatics. In
essence, the tracer refuses to inline functions which contain loops, since
it is normally better to trace them separately. Languages like Python can trivially
detect most loops statically (e.g.~the use of a \texttt{while} loop
in a function). Prolog loops, in contrast, are tail-recursive calls,
which, due to Prolog's highly dynamic nature, can often only
be detected reliably at run-time. By default,
the tracer therefore considers every call to Prolog as a potential loop
and does not inline any of them.

Our solution is simple, pragmatic, and can easily be applied to similar languages.
Rather than trying to devise complex heuristics for determining loops in Prolog,
we always inline a fixed number of Prolog reduction steps into Python.
If the Prolog predicate finishes execution within
that limit, then it is fully inlined. If it exceeds the limit, a call to the
rest of the predicate is emitted, stopping inlining at that point.
This means that if the Prolog predicate being inlined does contain a loop,
then part of it is unrolled into the machine
code for the outer Python function. Remaining loop iterations (if
any) are executed via an explicitly inserted call to the rest of the predicate.
This approach allows small predicates to benefit from inlining, while preventing
large predicates from causing unpalatable slow-downs. Ultimately the effect is
similar to the meta-tracer's normal inlining behaviour in mono-language
implementations like PyPy.

\subsection{\jythog}
\label{tuprolog}

\jythog composes two languages atop the JVM:
Jython\footnote{\url{http://jython.org/}} and
tuProlog~\cite{denti01tuprolog}. Jython is the only mature Python JVM
implementation and targets Python 2.7. We chose tuProlog since it is the most complete
embeddable JVM Prolog implementation we know of.\footnote{We also considered Jtrolog,
JLog, W-prolog, and GNU Prolog for Java.}~
Neither Jython nor tuProlog were altered in realising \jythog.

\section{Experimental Setup}
\label{sec:experiment}

This paper's main hypothesis is as follows:
\begin{description}
    \item[H1] Meta-tracing leads to faster interpreter composition than traditional approaches.
\end{description}
To address this hypothesis, we run a number of benchmarks across the four
compositions we created. The benchmarks are organised into two groups:
microbenchmarks, which
aim to test a single aspect of cross-language composition; and larger
benchmarks, which aim to test several aspects at once.

Running such benchmarks in a fair, consistent manner is tricky:
bias -- often unintentional and unnoted -- is rife.
This is particularly true for virtual machines with complex JIT compilers and
garbage collectors. These components bring in many sources of non-determinism,
many of them related to code and data layout differences in the JIT-compiled
code, which can lead to significant performance effects via different caching
behaviour~\cite{gu06relative,georges07statistically}. The fact that we combine
VMs that each have their individual effects exacerbates the issue.

We therefore base
our experimental methodology on the statistically rigorous \kalibera
method~\cite{kalibera13rigorous}. Since, to the best of our knowledge, we are the
first people apart from \kalibera to do so, we assume that most of our readers
are unfamiliar with this method, and give a gentle introduction in Section~\ref{sec:kaljo}.

\subsection{Microbenchmarks}
\label{sec:microbenchmarks}

We have designed six microbenchmarks that measure the costs of specific activities
when crossing language boundaries. Each benchmark consists of a `caller' function and
`callee' function:

\begin{description}
    \item[SmallFunc] The caller passes an integer to a tiny callee which
        returns the integer, incremented by 1.
    \item[L1A0R] Both the caller and the callee functions are loops. The
        callee receives a single integer argument.
    \item[L1A1R] Both the caller and callee are loops. The
        callee receives a single integer argument and returns a single
        integer result.
    \item[NdL1A1R] The caller invokes the callee in a
        loop. The callee produces more than one integer result (by
        returning an iterator in Python, and leaving a choice point in Prolog).
        The caller asks for all of the results (with a \texttt{for} loop in
        Python, and with a failure-driven loop in Prolog).
    \item[Lists] The callee produces a list, and the caller consumes
        it (e.g.~iterates over all its elements). The lists are converted between Prolog linked lists and Python
        array-based lists when passing the language barrier.
    \item[TCons] The caller walks a linked list which is produced by
        the callee using Prolog terms.
\end{description}

\noindent Each of the microbenchmarks comes in three variants:

\begin{description}
    \item[Python] Both caller and callee are Python code.
    \item[Prolog] Both caller and callee are Prolog code.
    \item[Python $\rightarrow$ Prolog] The caller is Python code, the callee is Prolog code.
\end{description}

\noindent The first two variants give us a baseline, while the latter
measures cross-language calling.  Whilst we have created
Prolog $\rightarrow$ Python variants of the benchmarks, only \unipycation can
execute them (see Section~\ref{swithon suckage}) so we do not consider them
further.

\subsection{Larger Benchmarks}

The limitations of microbenchmarks are well known. Since there are no
pre-existing \unipycation programs, we have thus created three (relatively)
larger benchmarks, based on an educated guess about the style of programs that real users
might write. While we would have liked to create a larger set of bigger
benchmarks, we are somewhat constrained by inherent limitations in \jythog and
\swithon, which accept a narrower range of programs than \unipycation itself.
However, while we do not claim that these benchmarks are definitely
representative of programs that other people may find \unipycation useful for,
they do enable us to examine a different class of behaviours than
the microbenchmarks.

Each of the benchmarks has two variants: Prolog only; and composed Python / Prolog.
The composed benchmarks can be broadly split into two: those that cross the
language barrier frequently (SAT model enumeration) and those that spend
most of their time in Prolog (Connect 4 and the Tube benchmark).

\subsubsection{Connect 4}

This benchmark is based on the implementation described in
Section~\ref{sec:demonstration}. To make a sensible
benchmark, we removed the GUI, made the AI player deterministic, and pitted
the AI player against itself. We measure the time taken for a complete AI
versus AI game to execute.

\subsubsection{Tube Route Finder}

Given a start and end station, this benchmark finds all the shortest possible
routes\footnote{Accurately modelling the behaviour of new tourists, our program
fails to consider the costs of changing lines and chooses routes which visit
the fewest possible stations.} through an undirected graph representing
the London Underground.\footnote{The graph data comes from:
\url{http://commons.wikimedia.org/wiki/London_Underground_geographic_maps/CSV/}}
This is achieved by
an iterative deepening depth-first search, monotonically increasing the depth
until at least one solution is found; backtracking is used to enumerate all
shortest routes. In the composed variant, Python increases the
depth in a loop, with each iteration invoking Prolog to run a depth first
search. Despite this, the composed benchmark intentionally spends nearly
all of its time in Prolog and is thus a base-case of sorts for composed
programs which do not cross the language barrier frequently. We measure the time
taken to find routes between a predetermined pair of stations.

\subsubsection{SAT Model Enumeration}
\label{sec:sat_solver}

This benchmark exhaustively enumerates the solutions (models) to propositional
formulas
using Howe and King's small Prolog SAT
solver~\cite{howe10pearl}. As the models are enumerated, we count the number of
propositional variables assigned to true. The composed variant uses
Python for the outer model enumeration loop.
We measure the time taken to count the true literals appearing in the models of
a SAT instance borrowed from the SATLIB suite~\cite{hoos00satlib}.

Note that since tuProlog does not support coroutines, this benchmark cannot be
run under \jythog.

\subsection{\kalibera Method}
\label{sec:kaljo}

Benchmarks often exhibit non-deterministic behaviour which goes unnoticed or
unexplained, distorting results and our perceptions of them. In computing, the
default response has mostly been to ignore the issue and hope for the
best~\cite{vitek11repeatability}.
To help make our benchmarks as fair and representative as reasonably possible,
we have adopted the more statistically rigorous \kalibera
method~\cite{kalibera13rigorous}. The basic aim of the method is to
determine a suitable number of experimental repetitions for the various levels
involved in benchmarking, whilst retaining a reasonable level of precision. It
also gives a mathematically sound method for computing relative speed-ups of
benchmarks, including confidence intervals. We now give a brief
overview of the method and detail how we applied it to our benchmarks.

The method first requires us to consider the levels involved in the
experiment. At each of the levels there is an opportunity to repeat
experimentation. Typical levels for a VM-based language (from low to high)
might be:
the \emph{iterations level}, sequential repetitions within a process; the
\emph{executions level}, repetitions of the iteration level, each time spawning
a fresh process;
and the \emph{VMs level}, repetitions of compiling the VM. Multiple levels need to be
considered because benchmarks can be non-deterministic at one level, but not at
another. For example, benchmarks which are non-deterministic in their
initialisation code will exhibit variation at the executions level but not the
iterations level. Cursory inspection showed that for our experiments there was no
meaningful variation at the VM level, so we concentrated on the iterations and
executions levels.

Having identified the appropriate levels, a \emph{dimensioning} run of the experiment is
performed with many repetitions to determine each benchmark's
variance at each level.
The ordered measurements are first used to determine if each benchmark reaches
an \emph{initialised state}: the point at which subsequent runs
are no longer subject to start-up costs (e.g.~JIT warmup). This is determined by
plotting a graph of sequential run times and manually determining when such a
state has been reached. The method also gives a process to determine if a benchmark
reaches an \emph{independent state}. If the benchmark reaches the
independent state, then we can be sure that the measurements taken for
subsequent iterations level runs are in no way correlated with previous
runs.  This is a useful property since it means that one can safely
take a sample of measurements at any point from the independent state
onwards. Since some experiments may never converge upon the
independent state, the methodology also provides a fallback method, whereby
a single consistent post-initialised iteration (in our case, the first) is taken.

For dimensioning, we used 10 runs at the iterations level and 10 at the executions level
-- i.e.~$10*10=100$ runs -- which was large enough to give useful data in
a somewhat reasonable time frame (about 3 weeks). To make our life
easier, for each VM we took the maximum initialisation sequence position (rather than
using a manually determined value for each VM / benchmark combination). We set
the initialisation state as iteration \#{}6 for all VMs except
for \swithon which, not using a JIT, reaches an initialised state at iteration \#{}2.
Since \jythog is so much slower than other implementations, we were unable to do 
a full dimensioning run in reasonable time; we thus used a subset of the benchmarks
to determine that it also reached the initialisation state at iteration \#{}6.

Given data from a dimensioning run, one can then compute the amount of variance
introduced at each level. To our surprise, the variation in our benchmarks
was consistently much higher at the executions level than the
iterations level. The method's heuristic therefore suggested that, since there was
little variation at the iterations level, we need only measure the time
taken for a
single post-initialisation iteration.  Since the methodology does not give a
method for choosing a number of repetitions for the highest level of
experimentation, we chose 30 repetitions for the
executions level. However, we were forced to lower this for \jythog, which
is rather slow (30 repetitions would lead to a running time of 3 months),
instead using 10 repetitions for \jythog's execution level.

With the repetition counts determined, we then ran the real experiment and
computed 99\% confidence intervals. As suggested by \kalibera, we used the
bootstrap re-sampling method with 10000 iterations to compute the confidence
intervals, which has the advantage that we are not required to make any
assumptions about the shape of the underlying distribution.

Our experiments were run on an otherwise idle system with a 4GHz Intel Core i7-4790K CPU
(with turbo mode disabled), 24 GiB RAM, and running Debian Linux 7.7 in
64 bit mode. All timings are wall-clock with a sub-microsecond resolution.

\subsection{Investigating conversion and the effect of meta-tracing}
\label{sec:more_experiments}

In addition to measuring the raw benchmarks, we perform two additional experiments designed to help us address two additional hypotheses:
\begin{description}
    \item[H2] Cross-language inlining is a significant part of meta-tracing's benefits.
    \item[H3] Meta-tracing reduces data conversion costs.
\end{description}

To test hypothesis H2, we created a variant of \unipycation that completely
disables inlining across the Python/Prolog boundary and ran all benchmarks on
that variant.

To test hypothesis H3, we created another benchmark variant
\pyprolognc which measures how fast the benchmarks run
with zero-cost conversions. For each benchmark we hand-crafted two versions: one
that computes and records the arguments and results for every cross-interpreter
call; and one that uses the precomputed arguments and results. The latter
variant is the one for which we measure times.

\section{Experimental Results}
\label{sec:results}

Tables~\ref{tab:crossings}, \ref{tab:micro}, and \ref{tab:larger}
show the raw results from running the benchmarks against the four composed VMs.
Table~\ref{tab:crossings} shows how often each benchmark crosses the language barrier
and how many objects are converted in those crossings;
Tables~\ref{tab:micro} and \ref{tab:larger} show benchmark timings in absolute seconds.

Tables~\ref{tab:micro-speedup} and \ref{tab:larger-speedup} show two subtly
different things.  First, the relative costs of moving from a mono-language to
a composed VM; informally, we say the composed VM is efficient if the relative
cost of the composed variant is `close' to the mono-language variant. Second,
the slow-down of using a composed VM relative to \unipycation.

In the rest of this section, we analyse these results in detail and share our insights into them.

\begin{table}[t]
\begin{center}
\begin{tabular}{p{.5\textwidth}rr}
\toprule
Benchmark                & Crossings      & ~~~~Objs converted \\
\midrule
SmallFunc                &   $1\,000\,000$&   $5\,000\,000$ \\
L1A0R         &      $100\,000$&      $100\,000$ \\
L1A1R         &      $100\,000$&      $300\,000$ \\
NdL1A1R                  & $100\,300\,000$& $100\,300\,000$ \\
TCons        &      $100\,000$& $400\,400\,000$ \\
Lists                    &      $100\,000$& $100\,500\,000$ \\
\midrule
sat-models               &       $41\,505$&  $12\,579\,962$ \\
tube                     &           $558$&        $3\,984$ \\
connect4                 &           $126$&       $14\,889$ \\
\bottomrule
\end{tabular}

\caption{How often each of our benchmarks crosses the language barrier, and how
many objects are converted in those crossings.}
\label{tab:crossings}
\end{center}
\end{table}

\begin{table*}[h!]
\begin{center}
{\small
\begin{tabular}{llrlrlrl}
\toprule
\multicolumn{1}{c}{VM}&Benchmark
& \multicolumn{2}{c}{\emph{Python}}
& \multicolumn{2}{c}{\emph{Prolog}}
& \multicolumn{2}{c}{\emph{Python} $\rightarrow$ \emph{Prolog}}
\\
\midrule
\multirow{6}{*}{\swithon} & SmallFunc
& 0.933s & {\tiny$\pm 0.002$}
& 1.983s & {\tiny$\pm 0.011$}
& 196.000s & {\tiny$\pm 4.225$}
\\
 & L1A0R
& 3.150s & {\tiny$\pm 0.008$}
& 5.712s & {\tiny$\pm 0.035$}
& 6.895s & {\tiny$\pm 0.024$}
\\
 & L1A1R
& 4.820s & {\tiny$\pm 0.006$}
& 15.279s & {\tiny$\pm 0.054$}
& 16.378s & {\tiny$\pm 0.056$}
\\
 & NdL1A1R
& 6.637s & {\tiny$\pm 0.011$}
& 12.186s & {\tiny$\pm 0.129$}
& 427.917s & {\tiny$\pm 3.239$}
\\
 & TCons
& 237.952s & {\tiny$\pm 1.209$}
& 36.034s & {\tiny$\pm 0.100$}
& 1451.516s & {\tiny$\pm 7.143$}
\\
 & Lists
& 9.328s & {\tiny$\pm 0.009$}
& 19.463s & {\tiny$\pm 0.060$}
& 1379.922s & {\tiny$\pm 7.618$}
\\
\midrule
\multirow{6}{*}{\swipy} & SmallFunc
& 0.006s & {\tiny$\pm 0.001$}
& 1.986s & {\tiny$\pm 0.014$}
& 20.840s & {\tiny$\pm 0.403$}
\\
 & L1A0R
& 0.049s & {\tiny$\pm 0.002$}
& 5.714s & {\tiny$\pm 0.042$}
& 5.774s & {\tiny$\pm 0.019$}
\\
 & L1A1R
& 0.074s & {\tiny$\pm 0.003$}
& 15.232s & {\tiny$\pm 0.039$}
& 14.628s & {\tiny$\pm 0.048$}
\\
 & NdL1A1R
& 0.423s & {\tiny$\pm 0.010$}
& 12.171s & {\tiny$\pm 0.119$}
& 50.424s & {\tiny$\pm 0.529$}
\\
 & TCons
& 5.576s & {\tiny$\pm 0.032$}
& 36.040s & {\tiny$\pm 0.057$}
& 104.392s & {\tiny$\pm 1.284$}
\\
 & Lists
& 0.701s & {\tiny$\pm 0.003$}
& 19.472s & {\tiny$\pm 0.088$}
& 71.818s & {\tiny$\pm 0.341$}
\\
\midrule
\multirow{6}{*}{\unipycation} & SmallFunc
& 0.008s & {\tiny$\pm 0.000$}
& 0.048s & {\tiny$\pm 0.007$}
& 0.010s & {\tiny$\pm 0.001$}
\\
 & L1A0R
& 0.045s & {\tiny$\pm 0.002$}
& 0.047s & {\tiny$\pm 0.003$}
& 0.045s & {\tiny$\pm 0.001$}
\\
 & L1A1R
& 0.069s & {\tiny$\pm 0.003$}
& 0.071s & {\tiny$\pm 0.003$}
& 0.074s & {\tiny$\pm 0.003$}
\\
 & NdL1A1R
& 0.422s & {\tiny$\pm 0.000$}
& 0.442s & {\tiny$\pm 0.016$}
& 2.493s & {\tiny$\pm 0.019$}
\\
 & TCons
& 4.754s & {\tiny$\pm 0.035$}
& 2.143s & {\tiny$\pm 0.003$}
& 28.871s & {\tiny$\pm 0.449$}
\\
 & Lists
& 0.700s & {\tiny$\pm 0.002$}
& 1.252s & {\tiny$\pm 0.002$}
& 4.176s & {\tiny$\pm 0.012$}
\\
\midrule
\multirow{6}{*}{\jythog} & SmallFunc
& 0.700s & {\tiny$\pm 0.013$}
& 23.788s & {\tiny$\pm 0.080$}
& 313.299s & {\tiny$\pm 3.294$}
\\
 & L1A0R
& 1.104s & {\tiny$\pm 0.007$}
& 153.766s & {\tiny$\pm 1.685$}
& 150.828s & {\tiny$\pm 1.032$}
\\
 & L1A1R
& 1.707s & {\tiny$\pm 0.062$}
& 222.879s & {\tiny$\pm 0.620$}
& 220.224s & {\tiny$\pm 1.994$}
\\
 & NdL1A1R
& 5.524s & {\tiny$\pm 0.134$}
& 1880.981s & {\tiny$\pm 7.642$}
& 2000.043s & {\tiny$\pm 15.500$}
\\
 & TCons
& 544.376s & {\tiny$\pm 6.075$}
& 7845.380s & {\tiny$\pm 28.407$}
& 8001.564s & {\tiny$\pm 26.390$}
\\
 & Lists
& 5.114s & {\tiny$\pm 0.079$}
& 7043.844s & {\tiny$\pm 15.929$}
& 5266.634s & {\tiny$\pm 13.188$}
\\
\bottomrule
\end{tabular}
}

\caption{Absolute performance (in seconds) of the microbenchmarks under each composed VM.}
\label{tab:micro}
\end{center}
\end{table*}

\begin{table*}[h!]
\begin{center}
{\small
\begin{tabular}{llrlrlrl}
\toprule
\multicolumn{1}{c}{VM}&Benchmark
& \multicolumn{2}{c}{$\frac{\mbox{\emph{Python}}\rightarrow\mbox{\emph{Prolog}}}{\mbox{\emph{Python}}}$}
& \multicolumn{2}{c}{$\frac{\mbox{\emph{Python}}\rightarrow\mbox{\emph{Prolog}}}{\mbox{\emph{Prolog}}}$}
& \multicolumn{2}{c}{$\frac{\mbox{\emph{Python}}\rightarrow\mbox{\emph{Prolog}}}{\mbox{\unipycation}}$}
\\
\midrule
\multirow{6}{*}{\swithon} & SmallFunc
& 210.060$\times$ & {\tiny$\pm 4.465$}
& 98.840$\times$ & {\tiny$\pm 2.188$}
& 20433.776$\times$ & {\tiny$\pm 1381.309$}
\\
 & L1A0R
& 2.189$\times$ & {\tiny$\pm 0.009$}
& 1.207$\times$ & {\tiny$\pm 0.009$}
& 151.951$\times$ & {\tiny$\pm 3.853$}
\\
 & L1A1R
& 3.398$\times$ & {\tiny$\pm 0.012$}
& 1.072$\times$ & {\tiny$\pm 0.005$}
& 222.353$\times$ & {\tiny$\pm 10.662$}
\\
 & NdL1A1R
& 64.473$\times$ & {\tiny$\pm 0.493$}
& 35.114$\times$ & {\tiny$\pm 0.452$}
& 171.661$\times$ & {\tiny$\pm 1.860$}
\\
 & TCons
& 6.100$\times$ & {\tiny$\pm 0.044$}
& 40.281$\times$ & {\tiny$\pm 0.232$}
& 50.276$\times$ & {\tiny$\pm 0.811$}
\\
 & Lists
& 147.939$\times$ & {\tiny$\pm 0.842$}
& 70.901$\times$ & {\tiny$\pm 0.454$}
& 330.461$\times$ & {\tiny$\pm 2.112$}
\\
\midrule
\multirow{6}{*}{\swipy} & SmallFunc
& 3759.944$\times$ & {\tiny$\pm 535.177$}
& 10.494$\times$ & {\tiny$\pm 0.219$}
& 2172.673$\times$ & {\tiny$\pm 150.436$}
\\
 & L1A0R
& 118.906$\times$ & {\tiny$\pm 5.622$}
& 1.010$\times$ & {\tiny$\pm 0.008$}
& 127.247$\times$ & {\tiny$\pm 3.153$}
\\
 & L1A1R
& 197.594$\times$ & {\tiny$\pm 8.467$}
& 0.960$\times$ & {\tiny$\pm 0.004$}
& 198.598$\times$ & {\tiny$\pm 9.384$}
\\
 & NdL1A1R
& 119.137$\times$ & {\tiny$\pm 2.999$}
& 4.143$\times$ & {\tiny$\pm 0.062$}
& 20.228$\times$ & {\tiny$\pm 0.263$}
\\
 & TCons
& 18.721$\times$ & {\tiny$\pm 0.251$}
& 2.897$\times$ & {\tiny$\pm 0.035$}
& 3.616$\times$ & {\tiny$\pm 0.072$}
\\
 & Lists
& 102.497$\times$ & {\tiny$\pm 0.646$}
& 3.688$\times$ & {\tiny$\pm 0.025$}
& 17.199$\times$ & {\tiny$\pm 0.096$}
\\
\midrule
\multirow{6}{*}{\unipycation} & SmallFunc
& 1.276$\times$ & {\tiny$\pm 0.081$}
& 0.201$\times$ & {\tiny$\pm 0.038$}
&1.000$\times$ & ~~
\\
 & L1A0R
& 1.005$\times$ & {\tiny$\pm 0.053$}
& 0.957$\times$ & {\tiny$\pm 0.057$}
&1.000$\times$ & ~~
\\
 & L1A1R
& 1.072$\times$ & {\tiny$\pm 0.071$}
& 1.034$\times$ & {\tiny$\pm 0.069$}
&1.000$\times$ & ~~
\\
 & NdL1A1R
& 5.902$\times$ & {\tiny$\pm 0.044$}
& 5.635$\times$ & {\tiny$\pm 0.216$}
&1.000$\times$ & ~~
\\
 & TCons
& 6.073$\times$ & {\tiny$\pm 0.106$}
& 13.471$\times$ & {\tiny$\pm 0.208$}
&1.000$\times$ & ~~
\\
 & Lists
& 5.969$\times$ & {\tiny$\pm 0.025$}
& 3.335$\times$ & {\tiny$\pm 0.010$}
&1.000$\times$ & ~~
\\
\midrule
\multirow{6}{*}{\jythog} & SmallFunc
& 447.378$\times$ & {\tiny$\pm 9.737$}
& 13.170$\times$ & {\tiny$\pm 0.146$}
& 32662.702$\times$ & {\tiny$\pm 2186.768$}
\\
 & L1A0R
& 136.595$\times$ & {\tiny$\pm 1.257$}
& 0.981$\times$ & {\tiny$\pm 0.013$}
& 3324.147$\times$ & {\tiny$\pm 87.511$}
\\
 & L1A1R
& 128.989$\times$ & {\tiny$\pm 5.334$}
& 0.988$\times$ & {\tiny$\pm 0.009$}
& 2989.890$\times$ & {\tiny$\pm 140.681$}
\\
 & NdL1A1R
& 362.045$\times$ & {\tiny$\pm 9.169$}
& 1.063$\times$ & {\tiny$\pm 0.009$}
& 802.325$\times$ & {\tiny$\pm 8.918$}
\\
 & TCons
& 14.699$\times$ & {\tiny$\pm 0.178$}
& 1.020$\times$ & {\tiny$\pm 0.005$}
& 277.151$\times$ & {\tiny$\pm 4.256$}
\\
 & Lists
& 1029.866$\times$ & {\tiny$\pm 15.745$}
& 0.748$\times$ & {\tiny$\pm 0.003$}
& 1261.242$\times$ & {\tiny$\pm 4.698$}
\\
\bottomrule
\end{tabular}
}

\caption{Relative performance of the microbenchmarks under each composed VM.
$\frac{\mbox{\emph{Python}}\rightarrow\mbox{\emph{Prolog}}}{\mbox{\emph{Python}}}$ should be read as
``the performance of the composed VM relative to the performance of the Python
part of the composed VM in isolation'' (similarly when Prolog is the divisor).
The first two numeric columns therefore show how expensive it is to move from a
mono-language Python/Prolog program to a composed variant. The third numeric
column shows the slow-down of using a composed VM relative to the composed
version running on \unipycation.}
\label{tab:micro-speedup}
\end{center}
\end{table*}

\begin{table*}[h!]
\begin{center}
{\small
\begin{tabular}{llrlrl}
\toprule
\multicolumn{1}{c}{VM}&Benchmark
& \multicolumn{2}{c}{\emph{Prolog}}
& \multicolumn{2}{c}{\emph{Python} $\rightarrow$ \emph{Prolog}}
\\
\midrule
\multirow{3}{*}{\swithon} & sat-models
& 3.779s & {\tiny$\pm 0.090$}
& 91.205s & {\tiny$\pm 0.629$}
\\
 & tube
& 0.272s & {\tiny$\pm 0.001$}
& 0.286s & {\tiny$\pm 0.001$}
\\
 & connect4
& 14.172s & {\tiny$\pm 0.033$}
& 14.324s & {\tiny$\pm 0.036$}
\\
\midrule
\multirow{3}{*}{\swipy} & sat-models
& 3.148s & {\tiny$\pm 0.006$}
& 12.156s & {\tiny$\pm 0.208$}
\\
 & tube
& 0.263s & {\tiny$\pm 0.002$}
& 0.265s & {\tiny$\pm 0.001$}
\\
 & connect4
& 14.172s & {\tiny$\pm 0.035$}
& 14.243s & {\tiny$\pm 0.045$}
\\
\midrule
\multirow{3}{*}{\unipycation} & sat-models
& 1.958s & {\tiny$\pm 0.024$}
& 2.862s & {\tiny$\pm 0.021$}
\\
 & tube
& 2.348s & {\tiny$\pm 0.026$}
& 2.381s & {\tiny$\pm 0.036$}
\\
 & connect4
& 7.533s & {\tiny$\pm 0.047$}
& 10.783s & {\tiny$\pm 0.078$}
\\
\midrule
\multirow{3}{*}{\jythog} & sat-models
& n/a & ~~
& n/a & ~~
\\
 & tube
& 435.555s & {\tiny$\pm 1.644$}
& 411.496s & {\tiny$\pm 0.607$}
\\
 & connect4
& 278.352s & {\tiny$\pm 0.787$}
& 268.351s & {\tiny$\pm 1.606$}
\\
\bottomrule
\end{tabular}
}

\caption{Absolute performance (in seconds) of the larger benchmarks under each composed VM.}
\label{tab:larger}
\end{center}
\end{table*}

\begin{table*}[h!]
\begin{center}
{\small
\begin{tabular}{llrlrl}
\toprule
\multicolumn{1}{c}{VM}&Benchmark
& \multicolumn{2}{c}{$\frac{\mbox{\emph{Python}}\rightarrow\mbox{\emph{Prolog}}}{\mbox{\emph{Prolog}}}$}
& \multicolumn{2}{c}{$\frac{\mbox{\emph{Python}}\rightarrow\mbox{\emph{Prolog}}}{\mbox{\unipycation}}$}
\\
\midrule
\multirow{3}{*}{\swithon} & sat-models
& 24.137$\times$ & {\tiny$\pm 0.599$}
& 31.871$\times$ & {\tiny$\pm 0.310$}
\\
 & tube
& 1.052$\times$ & {\tiny$\pm 0.005$}
& 0.120$\times$ & {\tiny$\pm 0.002$}
\\
 & connect4
& 1.011$\times$ & {\tiny$\pm 0.004$}
& 1.328$\times$ & {\tiny$\pm 0.010$}
\\
\midrule
\multirow{3}{*}{\swipy} & sat-models
& 3.862$\times$ & {\tiny$\pm 0.065$}
& 4.248$\times$ & {\tiny$\pm 0.076$}
\\
 & tube
& 1.008$\times$ & {\tiny$\pm 0.009$}
& 0.111$\times$ & {\tiny$\pm 0.002$}
\\
 & connect4
& 1.005$\times$ & {\tiny$\pm 0.004$}
& 1.321$\times$ & {\tiny$\pm 0.010$}
\\
\midrule
\multirow{3}{*}{\unipycation} & sat-models
& 1.462$\times$ & {\tiny$\pm 0.021$}
&1.000$\times$ & ~~
\\
 & tube
& 1.014$\times$ & {\tiny$\pm 0.019$}
&1.000$\times$ & ~~
\\
 & connect4
& 1.431$\times$ & {\tiny$\pm 0.014$}
&1.000$\times$ & ~~
\\
\midrule
\multirow{3}{*}{\jythog} & sat-models
& n/a & ~~
& n/a & ~~
\\
 & tube
& 0.945$\times$ & {\tiny$\pm 0.004$}
& 172.798$\times$ & {\tiny$\pm 2.599$}
\\
 & connect4
& 0.964$\times$ & {\tiny$\pm 0.006$}
& 24.887$\times$ & {\tiny$\pm 0.234$}
\\
\bottomrule
\end{tabular}
}

\caption{Relative performance of the larger benchmarks under each composed VM. See the caption
for Table~\ref{tab:micro-speedup} for details.}
\label{tab:larger-speedup}
\end{center}
\end{table*}

\subsection{Microbenchmarks}
\label{sec:microbenchmark analysis}

The microbenchmarks outlined in Section~\ref{sec:microbenchmarks} are designed to measure
the costs of specific cross-language activities and enable us to understand the
behaviour of larger benchmarks. Looking at Table~\ref{tab:micro}, several things
are worthy of note. We start with the simplest observations, moving gradually
to more subtle details.

The timings of the mono-language Python benchmarks are
near-identical for \unipycation and \swipy since they both execute under PyPy;
Jython is somewhat slower; and CPython considerably
slower (bar TCons and NdL1A1R). These results are roughly consistent with other benchmarking of
Python VMs~\cite{bolz14impact}, though these microbenchmarks flatter Jython,
which would typically be somewhat closer to CPython in performance.

The mono-language Prolog benchmarks
follow a similar pattern. \swithon and \swipy are equivalent within the margin of error because both
use SWI Prolog; Pyrolog is at least an order of magnitude faster; and tuProlog is 1--2
orders of magnitude slower than SWI Prolog. These results are broadly consistent
with previous benchmarking involving these Prolog VMs~\cite{bolz10towards},
though one should bear in mind that these microbenchmarks disproportionately flatter
Pyrolog. As these results show, tuProlog is so slow that it is close to
unusable (see Section~\ref{tuprolog suckage} for discussion of this).

The first two numeric columns of table~\ref{tab:micro-speedup} show that composed programs generally run slower than
their mono-language cousins. There are two reasons for this. First, composed
programs must perform type conversion when crossing the language barrier:
as Table~\ref{tab:crossings} shows, the microbenchmarks cross the barrier
frequently and often convert a large number of objects (we break these costs
down in more detail in Section~\ref{sec:conversion_costs}). Second, composed
microbenchmarks tend to be dominated by the slower of the two language
implementations being used. In our case, the slower implementation is always
Prolog. Moving an appropriate chunk of code from the slower part of the
composition to the faster part can therefore outweigh the overheads
imposed by the language barrier.\footnote{This same observation is the basis for
a standard style of performance programming in languages like Python and Ruby,
where small chunks of code are moved to C.} In our
benchmarks, we can see this behaviour clearly in the SmallFunc benchmark on
\unipycation: function calls in PyPy are much faster than Pyrolog's equivalents, because
the latter has additional machinery for unification, backtracking and non-determinism that cannot
always be optimised away.

The final numeric column in table~\ref{tab:micro-speedup} shows clearly that \unipycation is hugely faster on
these microbenchmarks than any other composed VM. In general it is at least one order of
magnitude faster, often two, and occasionally three or four orders of magnitude faster.
\unipycation benefits hugely from its ability to trace and optimise across language boundaries
(we examine this in more detail in Section~\ref{sec:noxtrace}). An exceptional case is the TCons
benchmark on \swipy, which manages a mere factor of 3.6 slowdown over
\unipycation. This is
because most of the cost of this benchmark is the conversion of a great many Prolog terms
(see Table~\ref{tab:crossings}) into Python lists, which is handled in the
high-level \texttt{uni} module. Since the \texttt{uni} module is a pure Python
module, both \swipy and \unipycation use PyPy for this aspect, and thus
\unipycation has less of an advantage.

\subsection{Larger Benchmarks}

With the results of the microbenchmarks in mind, two of the three larger
benchmarks in Tables~\ref{tab:larger} and \ref{tab:larger-speedup} are
relatively easily understood. \label{why_swi_is_better_for_tube}
The tube benchmark spends nearly all of its time
in Prolog (as Table~\ref{tab:crossings} shows, it makes few crossings) and thus
its results are dominated by the Prolog language implementation. In this case,
SWI Prolog considerably outperforms Pyrolog. The SAT models benchmark, in
contrast, crosses the language barrier frequently, passing large amounts of data
across. \unipycation thus outperforms
the other composed VMs by at least a factor of 4.

The Connect 4 benchmark is more
interesting. It crosses the language barrier relatively infrequently but, as
Table~\ref{tab:crossings} shows, passes a moderate amount of information (the game
state) in each crossing. The amount of data converted in the
\texttt{uni} module is small enough that substituting CPython for PyPy makes
little difference to performance (i.e.~\swithon and \swipy perform
near-identically). Table~\ref{tab:larger} shows that, in the pure
Prolog variant, Pyrolog runs this benchmark roughly twice as fast as SWI Prolog.
However, running it as a composed program on \unipycation leads to a surprisingly
large slowdown of around 40\%{}. Such a large difference cannot be explained by the
relatively small amount of data passed around. Manual inspection of the
traces compiled by the JIT reveals the cause to be the interaction between
two important Prolog predicates: \texttt{alphabeta}, which computes the AI's
next move; and \texttt{has\_won}, which decides if either player has won the
game.
In the pure Prolog variant, these two
predicates are optimised together. In the composed variant, the loop that calls
them is in Python, requiring a translation of the data from
\texttt{alphabeta} into Python and then back to Prolog to be passed to
\texttt{has\_won}. The presence of the conversion code prevents the two
Prolog predicates being optimised together, accounting for most
of the slowdown.

\jythog's results are interesting as the cross-language benchmark variants run
faster than the mono-language variants. In this case, tuProlog is so slow that
moving code into Jython can often be a net win. This emphasises 
Section~\ref{sec:microbenchmark analysis}'s observation that the language crossing
overhead can be smaller than the difference in performance between two language
implementations.

\subsection{The Effects of Cross-Language Tracing}
\label{sec:noxtrace}

\begin{table*}[t!]
\begin{center}
{\small
\begin{tabular}{llrlrlrl}
\toprule
\multicolumn{1}{c}{VM}&Benchmark
& \multicolumn{2}{c}{\emph{Python}}
& \multicolumn{2}{c}{\emph{Prolog}}
& \multicolumn{2}{c}{\emph{Python} $\rightarrow$ \emph{Prolog}}
\\
\midrule
\multirow{9}{*}{\unipycation NoXTrace} & SmallFunc
& 0.007s & {\tiny$\pm 0.001$}
& 0.049s & {\tiny$\pm 0.007$}
& 5.390s & {\tiny$\pm 0.013$}
\\
 & L1A0R
& 0.047s & {\tiny$\pm 0.002$}
& 0.046s & {\tiny$\pm 0.002$}
& 0.170s & {\tiny$\pm 0.001$}
\\
 & L1A1R
& 0.070s & {\tiny$\pm 0.003$}
& 0.071s & {\tiny$\pm 0.003$}
& 0.207s & {\tiny$\pm 0.001$}
\\
 & NdL1A1R
& 0.423s & {\tiny$\pm 0.000$}
& 0.401s & {\tiny$\pm 0.023$}
& 51.100s & {\tiny$\pm 0.050$}
\\
 & TCons
& 4.969s & {\tiny$\pm 0.020$}
& 2.144s & {\tiny$\pm 0.003$}
& 28.996s & {\tiny$\pm 0.092$}
\\
 & Lists
& 0.710s & {\tiny$\pm 0.004$}
& 1.250s & {\tiny$\pm 0.001$}
& 4.517s & {\tiny$\pm 0.009$}
\\
 & sat-models
& n/a & ~~
& 1.953s & {\tiny$\pm 0.017$}
& 2.523s & {\tiny$\pm 0.017$}
\\
 & tube
& n/a & ~~
& 2.363s & {\tiny$\pm 0.025$}
& 2.374s & {\tiny$\pm 0.019$}
\\
 & connect4
& n/a & ~~
& 7.396s & {\tiny$\pm 0.034$}
& 10.610s & {\tiny$\pm 0.114$}
\\
\bottomrule
\end{tabular}
}

\caption{Absolute performance of \unipycation with cross-language tracing turned
off.}
\label{tab:noinlining}
\end{center}
\end{table*}

\begin{table*}[t!]
\begin{center}
{\small
\begin{tabular}{llrlrlrl}
\toprule
\multicolumn{1}{c}{VM}&Benchmark
& \multicolumn{2}{c}{$\frac{\mbox{\emph{Python}}\rightarrow\mbox{\emph{Prolog}}}{\mbox{\emph{Python}}}$}
& \multicolumn{2}{c}{$\frac{\mbox{\emph{Python}}\rightarrow\mbox{\emph{Prolog}}}{\mbox{\emph{Prolog}}}$}
& \multicolumn{2}{c}{$\frac{\mbox{\emph{Python}}\rightarrow\mbox{\emph{Prolog}}}{\mbox{\unipycation}}$}
\\
\midrule
\multirow{9}{*}{\unipycation NoXTrace} & SmallFunc
& 798.291$\times$ & {\tiny$\pm 80.406$}
& 109.019$\times$ & {\tiny$\pm 16.048$}
& 561.895$\times$ & {\tiny$\pm 36.223$}
\\
 & L1A0R
& 3.657$\times$ & {\tiny$\pm 0.199$}
& 3.688$\times$ & {\tiny$\pm 0.184$}
& 3.756$\times$ & {\tiny$\pm 0.096$}
\\
 & L1A1R
& 2.961$\times$ & {\tiny$\pm 0.143$}
& 2.900$\times$ & {\tiny$\pm 0.143$}
& 2.814$\times$ & {\tiny$\pm 0.136$}
\\
 & NdL1A1R
& 120.903$\times$ & {\tiny$\pm 0.147$}
& 127.359$\times$ & {\tiny$\pm 7.736$}
& 20.499$\times$ & {\tiny$\pm 0.157$}
\\
 & TCons
& 5.836$\times$ & {\tiny$\pm 0.030$}
& 13.524$\times$ & {\tiny$\pm 0.046$}
& 1.004$\times$ & {\tiny$\pm 0.016$}
\\
 & Lists
& 6.358$\times$ & {\tiny$\pm 0.041$}
& 3.615$\times$ & {\tiny$\pm 0.008$}
& 1.082$\times$ & {\tiny$\pm 0.004$}
\\
 & sat-models
& n/a & ~~
& 1.292$\times$ & {\tiny$\pm 0.014$}
& 0.882$\times$ & {\tiny$\pm 0.009$}
\\
 & tube
& n/a & ~~
& 1.005$\times$ & {\tiny$\pm 0.014$}
& 0.997$\times$ & {\tiny$\pm 0.018$}
\\
 & connect4
& n/a & ~~
& 1.434$\times$ & {\tiny$\pm 0.017$}
& 0.984$\times$ & {\tiny$\pm 0.013$}
\\
\bottomrule
\end{tabular}
}

\caption{Relative performance of \unipycation with and without cross-language
tracing. See the caption for Table~\ref{tab:micro-speedup} for details.}
\label{tab:noinlining-speedup}
\end{center}
\end{table*}

The final numeric columns of Tables~\ref{tab:micro-speedup} and
\ref{tab:larger-speedup} compare the relative performance of the different
styles of composed VMs. Here, the outcome is clear (with the exception of the
tube benchmark, which rarely
crosses the language boundary): \unipycation outperforms the other styles of VM composition. This is
particularly interesting given that \swipy uses PyPy just as \unipycation does.
An important question therefore is whether the difference can be solely
explained by Pyrolog outperforming SWI Prolog. Looking at Tables~\ref{tab:micro}
and \ref{tab:larger}, we can see from the mono-language Prolog benchmarks that
this is not the case. Take, for example, the SmallFunc benchmark where Pyrolog
is about 42 times faster than SWI Prolog; in the composed variant of the
benchmark \unipycation is over 2000 times faster than \swipy. There must,
therefore, be one or more other factors which explain this.

Hypothesis H2 captures the intuition that cross-language inlining is likely
to explain this speedup. Tables~\ref{tab:noinlining} and
\ref{tab:noinlining-speedup} show the effect of turning off cross-language
inlining. Looking at
Table~\ref{tab:noinlining-speedup} we can clearly see that cross-language
tracing effects vary considerably depending on the benchmark. Most
of the microbenchmarks benefit from cross-language inlining, sometimes
substantially. For those microbenchmarks such as
SmallFunc and NdL1A1R that cross the language barrier frequently and call
functions that are small enough to be inlined by the meta-tracer,
cross-language tracing is highly effective. The effect is magnified for
SmallFunc because, unlike NdL1A1R, it generates a single solution
which can be highly optimised.

The effect with the larger benchmarks is rather different. Once we
account for confidence intervals, the effect of cross-language inlining on the
tube benchmark is inconclusive, connect4 slows down microscopically, and
sat-models slows down noticeably. At its root, sat-models is a
highly data-dependent benchmark, and some examples cause Pyrolog to generate
a large number of traces (around 600). On its own, this is not necessarily a problem---the
pure Prolog version of \unipycation is the fastest implementation of that
benchmark. However, these traces suffer from \emph{tail duplication}, a
common problem with tracing JIT compilers~\cite{gal09trace}, where
many traces have identical or near-identical code at their end. In the
case of sat-models, these tails are identical, and come from \unipycation (e.g.~for data
conversion) and PyPy (i.e.~the Python code that follows the call to Prolog).
Each of these needless tails causes around 1KiB of extra machine code per trace,
which is sufficient to cause cache misses, and thus largely accounts for the
slight slowdown. It is possible
that modifying Pyrolog with the tail call identifying techniques from
Pycket~\cite{baumann_et_al__pycket_a_tracing_jit_for_a_functional_language}
might reduce tail duplication in such examples.

While our data shows that hypothesis H2 sometimes explains \unipycation's
good performance, we consider our data insufficient to validate or refute hypothesis H2
in general.

\subsection{The Cost of Data Conversion}
\label{sec:conversion_costs}

\begin{table*}[t!]
\begin{center}
{\small
\begin{tabular}{llrlrl}
\toprule
\multicolumn{1}{c}{VM}&Benchmark
& \multicolumn{2}{c}{\emph{Python} $\overset{nc}{\rightarrow}$ \emph{Prolog}}
& \multicolumn{2}{c}{$\frac{\mbox{\emph{Python}}\rightarrow\mbox{\emph{Prolog}}}{\mbox{\emph{Python}}\overset{nc}{\rightarrow}\mbox{\emph{Prolog}}}$}
\\
\midrule
\multirow{9}{*}{\swithon} & SmallFunc
& 125.273s & {\tiny$\pm 4.514$}
& 1.565$\times$ & {\tiny$\pm 0.066$}
\\
 & L1A0R
& 6.901s & {\tiny$\pm 0.039$}
& 0.999$\times$ & {\tiny$\pm 0.007$}
\\
 & L1A1R
& 15.895s & {\tiny$\pm 0.135$}
& 1.030$\times$ & {\tiny$\pm 0.010$}
\\
 & NdL1A1R
& 229.359s & {\tiny$\pm 2.036$}
& 1.866$\times$ & {\tiny$\pm 0.023$}
\\
 & TCons
& 144.363s & {\tiny$\pm 0.457$}
& 10.055$\times$ & {\tiny$\pm 0.060$}
\\
 & Lists
& 10.988s & {\tiny$\pm 0.058$}
& 125.590$\times$ & {\tiny$\pm 0.980$}
\\
 & sat-models
& 5.715s & {\tiny$\pm 0.068$}
& 15.960$\times$ & {\tiny$\pm 0.226$}
\\
 & tube
& 0.260s & {\tiny$\pm 0.001$}
& 1.103$\times$ & {\tiny$\pm 0.006$}
\\
 & connect4
& 14.223s & {\tiny$\pm 0.051$}
& 1.007$\times$ & {\tiny$\pm 0.004$}
\\
\midrule
\multirow{9}{*}{\swipy} & SmallFunc
& 9.517s & {\tiny$\pm 0.139$}
& 2.190$\times$ & {\tiny$\pm 0.052$}
\\
 & L1A0R
& 5.791s & {\tiny$\pm 0.031$}
& 0.997$\times$ & {\tiny$\pm 0.006$}
\\
 & L1A1R
& 14.631s & {\tiny$\pm 0.048$}
& 1.000$\times$ & {\tiny$\pm 0.005$}
\\
 & NdL1A1R
& 18.864s & {\tiny$\pm 0.201$}
& 2.673$\times$ & {\tiny$\pm 0.043$}
\\
 & TCons
& 19.004s & {\tiny$\pm 0.071$}
& 5.493$\times$ & {\tiny$\pm 0.070$}
\\
 & Lists
& 7.389s & {\tiny$\pm 0.032$}
& 9.719$\times$ & {\tiny$\pm 0.063$}
\\
 & sat-models
& 3.041s & {\tiny$\pm 0.005$}
& 3.997$\times$ & {\tiny$\pm 0.066$}
\\
 & tube
& 0.257s & {\tiny$\pm 0.001$}
& 1.031$\times$ & {\tiny$\pm 0.006$}
\\
 & connect4
& 14.226s & {\tiny$\pm 0.042$}
& 1.001$\times$ & {\tiny$\pm 0.004$}
\\
\midrule
\multirow{9}{*}{\unipycation} & SmallFunc
& 0.006s & {\tiny$\pm 0.001$}
& 1.725$\times$ & {\tiny$\pm 0.257$}
\\
 & L1A0R
& 0.047s & {\tiny$\pm 0.002$}
& 0.969$\times$ & {\tiny$\pm 0.047$}
\\
 & L1A1R
& 0.072s & {\tiny$\pm 0.003$}
& 1.020$\times$ & {\tiny$\pm 0.068$}
\\
 & NdL1A1R
& 1.624s & {\tiny$\pm 0.010$}
& 1.535$\times$ & {\tiny$\pm 0.015$}
\\
 & TCons
& 4.800s & {\tiny$\pm 0.028$}
& 6.015$\times$ & {\tiny$\pm 0.102$}
\\
 & Lists
& 1.075s & {\tiny$\pm 0.002$}
& 3.884$\times$ & {\tiny$\pm 0.013$}
\\
 & sat-models
& 2.071s & {\tiny$\pm 0.086$}
& 1.382$\times$ & {\tiny$\pm 0.058$}
\\
 & tube
& 1.688s & {\tiny$\pm 0.024$}
& 1.411$\times$ & {\tiny$\pm 0.030$}
\\
 & connect4
& 10.773s & {\tiny$\pm 0.114$}
& 1.001$\times$ & {\tiny$\pm 0.012$}
\\
\midrule
\multirow{9}{*}{\unipycation NoXTrace} & SmallFunc
& 4.669s & {\tiny$\pm 0.019$}
& 1.154$\times$ & {\tiny$\pm 0.006$}
\\
 & L1A0R
& 0.171s & {\tiny$\pm 0.001$}
& 0.998$\times$ & {\tiny$\pm 0.006$}
\\
 & L1A1R
& 0.203s & {\tiny$\pm 0.001$}
& 1.020$\times$ & {\tiny$\pm 0.009$}
\\
 & NdL1A1R
& 45.280s & {\tiny$\pm 0.568$}
& 1.129$\times$ & {\tiny$\pm 0.014$}
\\
 & TCons
& 5.216s & {\tiny$\pm 0.025$}
& 5.559$\times$ & {\tiny$\pm 0.033$}
\\
 & Lists
& 1.328s & {\tiny$\pm 0.006$}
& 3.402$\times$ & {\tiny$\pm 0.018$}
\\
 & sat-models
& 1.985s & {\tiny$\pm 0.075$}
& 1.271$\times$ & {\tiny$\pm 0.049$}
\\
 & tube
& 1.680s & {\tiny$\pm 0.018$}
& 1.413$\times$ & {\tiny$\pm 0.020$}
\\
 & connect4
& 10.674s & {\tiny$\pm 0.056$}
& 0.994$\times$ & {\tiny$\pm 0.012$}
\\
\bottomrule
\end{tabular}
}

\caption{Absolute performance in seconds of all VMs with conversion costs
removed (left column) and relative performance of \pyprolog to \pyprolognc
(right column).}
\label{tab:noconvert}
\end{center}
\end{table*}

Table~\ref{tab:noconvert} shows the results from running the \pyprolognc
benchmarks (see Section~\ref{sec:more_experiments}).\footnote{Since
it would take two additional weeks to run, we did not run the \pyprolognc
benchmark on \jythog.} As one would expect, removing
conversion makes all benchmarks faster. The benchmarks that benefit the
most in terms of relative time reduction are TCons, Lists and
sat-models, which is consistent with the fact that these benchmarks convert a
large number of objects. NdL1A1R does not benefit as much; although many objects
are converted, they are simple integers rather than more complex structures.

The results also indirectly show that there are additional costs beyond data
conversion: additional function calls, predicate lookups, allocate additional
memory for other reasons etc. The \pyprolognc numbers are therefore still slower
than the pure Prolog or Python variants, despite the elimination of the conversion costs.
Of particular interest are \unipycation's results: although it performs better
without data conversion costs, it benefits much less than the other composed
variants, showing that meta-tracing is particularly effective at removing data
conversion costs in composed VMs. We believe that these results
validate hypothesis H3 fairly clearly.

\subsection{Disentangling Cross-language Inlining and Data Conversion}

By turning off cross-language inlining and using our modified zero-cost data
conversion benchmarks, we can understand their individual effects. The results
are shown in the \unipycation-NoXTrace parts of Table~\ref{tab:noconvert}.
As we would expect, the absolute times between
\unipycation and \unipycation-NoXTrace in Table~\ref{tab:noconvert} are often
substantially different; however, the relative times are generally fairly close.
For example, the
relative speed-up of the NdL1A1R benchmark on \unipycation is $1.535\times$
and for \unipycation-NoXTrace $1.129\times$. This shows that the benefits of
cross-language inlining and data conversion in meta-tracing are largely independent.

\subsection{Summary: The Cost of Moving to Composed VMs}

The first two numeric columns of Table~\ref{tab:micro-speedup} and first
numeric column of Table~\ref{tab:larger-speedup} give an
indication of the performance penalty users can expect to incur when moving from
a mono-language VM to a composed VM. This clearly shows that it is more costly
to move from Python to \unipycation than from Prolog to \unipycation. Put
another way, Python users who care only about performance may have to think more
carefully about such a move than Prolog users. There are several factors that
are likely to contribute to this: PyPy is generally faster than Pyrolog; it is
generally easier to optimise an imperative programming implementation such as
PyPy than a declarative implementation such as Pyrolog; and it is easier for an
end-user to write a Python program in such a way that an implementation can run
it efficiently relative to Prolog.

Fortunately, the results also show that the costs of moving from mono-language
PyPy or Pyrolog to composed \unipycation are not terrible.  Even for the worst-case
microbenchmark, the penalty for moving to \unipycation is an order of
magnitude, but more typically a factor of 1--6.  In the larger benchmarks,
the penalty for moving is at most a factor of 1.462, which many users may
consider an acceptable cost for the increase in expressiveness that composed
programming brings.

\section{Discussion}
\label{sec:discussion}

Section~\ref{sec:results} gave the quantitative results from our
experiment. In this section, we give our qualitative impressions, first of the
four different compositions, and then of the \kalibera method.

\subsection{\unipycation}

RPython's relatively high-level nature (e.g.~rich data types and garbage
collection) made much of \unipycation's development fairly easy. On the other
hand, compiling (or, in RPython terminology, \emph{translating}) RPython VMs is slow.
\unipycation takes the best part of an hour
to translate on a fast machine, a cost that is born anew for every single change
due to the lack of incremental compilation. A partial remedy used by the RPython
community is to unit test small fragments of functionality untranslated in a Python
interpreter. This inevitably gives little idea of the performance of a translated
VM, but can catch many errors quicker than running the translator (though some
errors are only likely to be found by the translator's type inference).

\subsection{\swithon and \swipy}
\label{swithon suckage}

Using CFFI eased the development of \swithon significantly.
We would not have enjoyed dealing with intricate
details such as CPython's reference counting and reference ownership
semantics. Since CFFI is portable across Python implementations, we were
able to run \swithon's glue code under PyPy without modification to
`create' \swipy. Despite the use of CFFI -- and despite the experience we accumulated having
implemented the other VMs -- \swithon and \swipy were without doubt the most
challenging compositions to create. There were two chief reasons for this.

First, we did not initially appreciate that SWI Prolog cannot interleave queries
within a single Prolog engine, so code such as the following fails to execute as
expected:
\begin{lstlisting}
it1 = e.db.f.iter(...)
it2 = e.db.g.iter(...)
s1 = it1.next()
\end{lstlisting}
\noindent After the second query is invoked, SWI Prolog does not allow the
first to be pumped for additional values. We eventually made line 3 raise an exception (\texttt{StaleSolutionIterator}) in \swithon\ / \swipy. One work around for this
would be to create a new Prolog engine for each query; however, SWI Prolog
requires a new thread for each engine, which would lead to unacceptably high
overheads.

Secondly, our unit testing approach, which worked well for \unipycation and
\jythog, was less effective for \swithon. Seemingly innocuous mistakes led to
the test suite crashing without useful debugging information or corrupted the
system such that the next test (which in isolation, would pass) then failed.

\subsection{\jythog}
\label{sec:discussjythog}

Since Jython can seamlessly
call Java from Python, implementing glue code for \jythog was fairly trivial.
However, \jythog's poor performance highlights the semantic mismatch
problem~\cite{bolz14impact}. VMs can't optimise every possible program;
instead, they optimise common idioms in the family of languages they
expect to encounter. Both Jython and tuProlog fall some way outside what current JVMs
can best optimise; \jythog exacerbates these issues, and performance suffers greatly.
We suggest that language composition on existing VMs such as the JVM is very
hard to optimise, unless all languages involved fall within the boundaries
of the expected language model.

\subsection{Effort Levels}

Reporting precise levels of effort for each composed VM is impractical. However,
we believe that, roughly speaking, each composition took about one person month to
implement; and that designing and implementing the \texttt{uni} interface took
roughly one person month as well (we count \swithon and \swipy as one composition in this regard).
\unipycation was implemented first; then \jythog; and then
\swithon (and therefore \swipy). We thus had the ability to learn from our
experiences along this journey, which perhaps made \jythog and \swithon slightly
easier to implement (though \swithon's proneness to segfault in unexpected ways
caused the most hair-pulling). The main lesson we drew from this was that while the
style of composition chosen has a big effect on benchmark performance, it does
not significantly effect how much effort is needed to create the composition.

\subsection{The \kalibera Method}

The \kalibera method is not yet commonplace and so we believe readers may gain
some insights from our experiences of it. The first thing to note is that the
method is not a prescriptive sequence of steps, but instead is a cookbook of
related techniques and suggestions. While some aspects are clearly more
important than others, there is plenty of scope for variation. It is also
more obviously aimed at long-running, rather than micro, benchmarks.

The method forced us to think about the multiple levels involved in executing a
benchmark. Previously, we had tended to benchmark either the iterations or the
executions level. We already knew that we should discard results prior to the
initialisation state, but the dimensioning run made it clear that the
variation at the executions level was much bigger than at the iterations
level. The method allowed us to understand these
issues, and to concentrate the experiment's time on the level(s) where
the variation was the greatest.

We deviated from the method by not applying separate initialisation state counts
to each benchmark. Instead, we took the maximum and applied that to all
benchmarks on a per-VM basis. This very slightly increased the time benchmarks took to execute but
made configuration easier, and has no detrimental effect on the final results. We consider
this a reasonable trade-off. We ignored the notion of independent state, as
the heuristics told us to perform just one (initialised) repetition at the iterations level.

One inherent limitation of the method is that it cannot choose repetition
counts for the highest level (for us, the executions level). In our case we chose 30 repetitions (and only 10 for \jythog) for this
level based on our intuition of the benchmarking in question.

One of the largest contributions of the method is that it brings together
some advanced statistics and makes them approachable. The method covers all
aspects of experimentation, from designing the experiment, to presenting the
results with confidence intervals (e.g.~Tables~\ref{tab:micro-speedup},
\ref{tab:larger-speedup} and \ref{tab:noinlining-speedup}). We
reimplemented the statistical aspects of the \kalibera
method as a Python package which can be freely downloaded.\footnote{\url{http://soft-dev.org/src/libkalibera/}}

\section{Threats to Validity}
\label{sec:threats}

As ever with microbenchmarks, one must not expect that they tell us much about
real-world programs. Our larger benchmarks are an attempt to gain a slightly
better idea of how benchmarks that cross the language barrier in different ways
perform. `Real' composed programs will be many times bigger than
our biggest benchmark and are likely to have different
performance characteristics.

\label{jythog suckage} The fact that the VM glue code which performs
high-level type conversions and so on is written in pure Python, rather
than in the language of the underlying language implementation, is a
potential worry. In particular, \swithon has no good way to dynamically
optimise this code (unlike \unipycation and \swipy). We are able to get a
good handle on the size of this effect through the \pyprolognc data from
Section~\ref{sec:conversion_costs}. By reducing all data conversion costs to
zero, we emulate an extreme version of what would happen if the glue code
was rewritten in C. In other words, we can use this as a good proxy for
answering the question `would \swithon outperform \unipycation if it's glue code
wasn't written in Python?' As the numbers clearly show, even when data
conversions costs are removed, \unipycation achieves significantly better
performance in all but the tube benchmark, which has no relevance from the
perspective of data conversions (see Page~\pageref{why_swi_is_better_for_tube}).
This is strong evidence that our overall conclusions would remain the
same even if \swithon's glue code were to be rewritten.

\label{tuprolog suckage} Although the same argument could be extended to \jythog,
it is largely irrelevant: tuProlog is so slow that it dominates
all benchmarks it is a part of. We stuck with tuProlog as it is the only JVM
Prolog we could find which is stable and easily embeddable. A different JVM
Prolog may lessen this effect, though the semantic mismatch problem (see
Section~\ref{sec:discussjythog}) suggests that it will always be difficult
to make Prolog perform well on the JVM.

\section{Related Work}
\label{sec:related}

The motivation for language composition dates back to the late
60s~\cite{cheatham69motivation}, though most of the early work was on extensible
languages (e.g.~\cite{irons70experience}); to the best of our knowledge such
work largely disappeared from view for many years, though there has been
occasional successor work
(e.g.~\cite{gregory85ametalanguage,cardelli93extensible}). We are not
aware of an extensive study of this early work and, unfortunately, the passing
of time has made it hard to relate much it to the current day---in some cases it
has been decades since the systems involved could be run.

The Domain Specific Language
(DSL) movement aims for a limited form of language composition. One part of the
movement (best represented by~\cite{hudak96building}) sees DSLs as a specific
way of using a language's existing syntax and semantics, and is of little
relevance to this paper. The other part of the movement aims to actively extend
a language's syntax and semantics. This is normally achieved either by
source-to-source translation, or by macro-esque compilation of an embedded DSL
into a host language. Translating one language into another generally leads to
semantic mismatches which cause poor performance (see
e.g.~\cite{tratt08domainspecific}), analogous to \jythog. However, it is perfectly possible to
implement DSLs as interpreters, and our results may give useful suggestions
to performance-concious DSL authors as to which approach may be most suitable
for them.

Arguably the most common approach to composing languages is through
an FFI (Foreign Function Interface). In most operating systems, C or C++ are
the de facto standard languages, and most other languages thus provide an FFI to
them. Traditional FFIs can not inline across languages which, as our results show,
often compromises performance.

Another approach to language composition is to glue together VMs running in
different processes through CORBA and its ilk
(e.g.~\cite{janssen94interlanguage}). However such approaches inevitably have to
take a `lowest-possible common denominator' approach, even when all parts of the
composition could communicate at a higher-level~\cite{kaplan98idls}, making
programming awkward and inefficient. The approaches we use in this paper impose
no such requirement on the composition, which can interoperate at whatever level
makes sense. Furthermore, our intra-process approaches can communicate without
the substantial overhead of inter-process communication (often 4--6 orders of
magnitude difference, even with the most efficient OS implementations). Because
of these issues, inter-process composition has never seen any meaningful use for
fine-grained composition, although it is used to glue together large components
in a coarse-grained fashion.

Semantic mismatches make it difficult to create performant language compositions
atop a single VM. While Java programs on HotSpot have excellent performance,
other languages (e.g.~Python) on HotSpot often run slower than simple C-based
interpreters~\cite{bolz14impact,sarimbekov13characteriustics}. While better VM extensions
(e.g.~\texttt{invokedynamic}) or compiler alterations
(e.g.~\cite{ishizaki12adding}) can improve performance, the results still lag
some way behind their tracing equivalents~\cite{castanos12onthebenefits}. Our
experience with \unipycation shows that this result holds true in the face of
language composition too.

Similar approaches to \unipycation exist at the language-level: Smalltalk
and SOUL (a Prolog-like logic programming language)
\cite{gybels03soul,dhondt04seamless}; Java and
tuProlog~\cite{denti01tuprolog}; Icon and Prolog~\cite{lapaime86logicon};
and Lisp and Prolog in LOGLISP~\cite{robinson80loglisp}.  Most of these
compositions have a similar cross-language API to \unipycation. For example,
SOUL predicates are mapped to message sends, and SOUL's multiple solutions
are mapped to collections (albeit not lazily). Note however that such
approaches simply embed a (slow) Prolog interpreter written in the host
language, leading to much worse performance than \unipycation's cross-language JIT compiler.
LOGLISP is the outlier, extending Lisp's primitives with Prolog-like features.
Due to LOGLISP's age, we have little idea about its performance characteristics.

\section{Conclusions}
\label{sec:conclusion}

In this paper we aimed to further our understanding of which style of composition
leads to the best run-time performance. By composing Python and Prolog, we
hope to have chosen a difficult scenario that gives an indication of the worst-case
in terms of performance and engineering effort.  Although we must be careful not to
over-generalise our experiences, our results are fairly clear and validate
hypothesis H1. Meta-tracing led
to a composed VM which at least holds its own with, and generally beats significantly, other
styles of composition. The composition was no more difficult to implement
than the next fastest style of composition, which
required interfacing at the C level.

Our current assumption is that this pattern is likely to be repeated for most
languages which are naturally suited to meta-tracing. However, languages which
block cross-language inlining will perform poorly. Fortunately, our experiences
with Pyrolog suggest that interpreters for such languages can be easily tweaked
such that meta-tracing can inline across the language boundary. An open
question is how our results may apply to languages less suited to meta-tracing.
For example, in isolation, a traditional static C compiler is likely to beat a meta-tracing
C interpreter in most cases. As our results have shown, the cost of repeatedly
crossing the language boundary can outweigh the performance characteristics of
individual interpreters. A validation of this can be seen
in~\cite{grimmer15dynamically} which uses dynamic partial-evaluation of
self-optimising interpreters -- an approach similar in intent to meta-tracing --
to show that composing a C interpreter with Ruby can give good results. This
leads us to suggest that perhaps the most important factor in determining which
interpreter composition should be used is the frequency with which one expects
users to cross the language boundary: if the answer is `not often', then the
implementation choice may not be hugely important; if `often', then meta-tracing
(or a similar approach) may well be the best choice, irrespective of the
individual languages in the composition.

Finally, while our work is pitched in the context of language
composition, much of the results and insights will be applicable to cross-language
libraries. This is relatively common-place on the JVM (e.g.~Scala calling Java), and
in many dynamically typed languages (e.g.~Ruby calling C). Our results give a good understanding
of how moving to different language implementation styles effects performance in
such cases.\\[10pt]

\noindent\textbf{Acknowledgements:} This research was funded by the EPSRC \emph{Cooler} grant EP/K01790X/1. We
thank Jim Baker, Tomas Kalibera, and Jan Wielemaker for help. Martin Berger and
Samuele Pedroni gave us insightful feedback on early drafts of this paper.

\bibliography{bib}

\begin{thebibliography}{10}
\providecommand{\url}[1]{\texttt{#1}}
\providecommand{\urlprefix}{URL }

\bibitem{bala00dynamo}
Bala, V., Duesterwald, E., Banerjia, S.: Dynamo: A transparent dynamic
  optimization system. In: PLDI. pp. 1--12 (Jun 2000)

\bibitem{barrett13unipycation}
Barrett, E., Bolz, C.F., Tratt, L.: {Unipycation}: A case study in
  cross-language tracing. In: VMIL. pp. 31--40 (Oct 2013)

\bibitem{baumann_et_al__pycket_a_tracing_jit_for_a_functional_language}
Baumann, S., Bolz, C.F., Hirschfeld, R., Kirilichev, V., Pape, T., Siek, J.,
  Tobin-Hochstadt, S.: Pycket: A tracing jit for a functional language. In:
  ICFP (2015), accepted for publication

\bibitem{bebenita10spur}
Bebenita, M., Brandner, F., Fahndrich, M., Logozzo, F., Schulte, W., Tillmann,
  N., Venter, H.: {SPUR}: A trace-based {JIT} compiler for {CIL}. In: OOPSLA.
  pp. 708--725 (Mar 2010)

\bibitem{bolz11allocation}
Bolz, C.F., Cuni, A., Fijałkowski, M., Leuschel, M., Pedroni, S., Rigo, A.:
  Allocation removal by partial evaluation in a tracing {JIT}. PEPM pp. 43--52
  (Jan 2011)

\bibitem{bolz09tracing}
Bolz, C.F., Cuni, A., Fijałkowski, M., Rigo, A.: Tracing the meta-level:
  {PyPy's} tracing {JIT} compiler. In: ICOOOLPS. pp. 18--25 (Jul 2009)

\bibitem{bolz10towards}
Bolz, C.F., Leuschel, M., Schneider, D.: Towards a jitting {VM} for {Prolog}
  execution. In: PPDP. pp. 99--108 (Jul 2010)

\bibitem{bolz14impact}
Bolz, C.F., Tratt, L.: The impact of meta-tracing on {VM} design and
  implementation. To appear~J.~SCICO  (2014)

\bibitem{bratko01prolog}
Bratko, I.: Prolog programming for artificial intelligence. Addison Wesley
  (2001)

\bibitem{cardelli93extensible}
Cardelli, L., Matthes, F., Abadi, M.: Extensible grammars for language
  specialization. In: Workshop on Database Programming Languages. pp. 11--31
  (Aug 1993)

\bibitem{castanos12onthebenefits}
Castanos, J., Edelsohn, D., Ishizaki, K., Nagpurkar, P., Nakatani, T.,
  Ogasawara, T., Wu, P.: On the benefits and pitfalls of extending a statically
  typed language {JIT} compiler for dynamic scripting languages. In: OOPSLA.
  pp. 195--212 (Oct 2012)

\bibitem{cheatham69motivation}
Cheatham, T.E.: Motivation for extensible languages. SIGPLAN  4(8),  45--49
  (Aug 1969)

\bibitem{denti01tuprolog}
Denti, E., Omicini, A., Ricci, A.: {tuProlog:} a light-weight {Prolog} for
  internet applications and infrastructures. In: PADL, vol. 1990, pp. 184--198
  (Mar 2001)

\bibitem{dhondt04seamless}
D'Hondt, M., Gybels, K., Jonckers, V.: Seamless integration of rule-based
  knowledge and object-oriented functionality with linguistic symbiosis. In:
  SAC. pp. 1328--1335 (Mar 2004)

\bibitem{edwards61alpha}
Edwards, D.J., Hart, T.P.: The alpha-beta heuristic. Tech. Rep. AIM-30, MIT
  (1961)

\bibitem{gal09trace}
Gal, A., Eich, B., Shaver, M., Anderson, D., Mandelin, D., Haghighat, M.R.,
  Kaplan, B., Hoare, G., Zbarsky, B., Orendorff, J., Ruderman, J., Smith, E.W.,
  Reitmaier, R., Bebenita, M., Chang, M., Franz, M.: Trace-based {Just-In-Time}
  type specialization for dynamic languages. In: PLDI. pp. 465--478 (Jun 2009)

\bibitem{gal06hotpathvm}
Gal, A., Probst, C.W., Franz, M.: {HotpathVM:} an effective {JIT} compiler for
  resource-constrained devices. In: VEE. pp. 144--153 (Jun 2006)

\bibitem{georges07statistically}
Georges, A., Buytaert, D., Eeckhout, L.: Statistically rigorous {Java}
  performance evaluation. SIGPLAN Not.  42(10),  57--76 (2007)

\bibitem{grimmer15dynamically}
Grimmer, M., Seaton, C., W\"urthinger, T., M\"ossenb\"ock, H.: Dynamically
  composing languages in a modular way: Supporting {C} extensions for dynamic
  languages (Mar 2015)

\bibitem{gu06relative}
Gu, D., Verbrugge, C., Gagnon, E.M.: Relative factors in performance analysis
  of {Java Virtual Machines}. In: VEE. pp. 111--121 (Jun 2006)

\bibitem{gybels03soul}
Gybels, K.: {SOUL} and {Smalltalk} -- just married: evolution of the
  interaction between a logic and an object-oriented language towards
  symbiosis. In: DP-COOL (Aug 2003)

\bibitem{hoos00satlib}
Hoos, H., St\"utzle, T.: {SATLIB}: An online resource for research on {SAT}.
  In: SAT 2000. pp. 283--292 (2000), i.~P.~Gent, H.~v.~Maaren, T.~Walsh,
  editors, SAT 2000. SATLIB is available online at \url{www.satlib.org}

\bibitem{howe10pearl}
Howe, J., King, A.: A pearl on {SAT} solving in {Prolog}. In: FLOPS. vol. 6009,
  pp. 165--174 (Apr 2010)

\bibitem{hudak96building}
Hudak, P.: Building domain-specific embedded languages. ACM Computing Surveys
  28(4) (Dec 1996)

\bibitem{ingalls97back}
Ingalls, D., Kaehler, T., Maloney, J., Wallace, S., Kay, A.: Back to the
  future: the story of {Squeak}, a practical {Smalltalk} written in itself. In:
  OOPSLA. pp. 318--326 (Oct 1997)

\bibitem{irons70experience}
Irons, E.T.: Experience with an extensible language. Communications of the ACM
  13(1),  31--40 (Jan 1970)

\bibitem{ishizaki12adding}
Ishizaki, K., Ogasawara, T., Castanos, J., Nagpurkar, P., Edelsohn, D.,
  Nakatani, T.: Adding dynamically-typed language support to a statically-typed
  language compiler: Performance evaluation, analysis, and tradeoffs. In: VEE.
  pp. 169--180 (Mar 2012)

\bibitem{janssen94interlanguage}
Janssen, B., Spreitzer, M.: {ILU}: Inter-language unification via object
  modules. In: Workshop on Multi-Language Object Models (Aug 1994)

\bibitem{gregory85ametalanguage}
Johnson, G.F., Fischer, C.N.: A meta-language and system for nonlocal
  incremental attribute evaluation in language-based editors. In: POPL. pp.
  141--151 (Jan 1985)

\bibitem{kalibera13rigorous}
Kalibera, T., Jones, R.: Rigorous benchmarking in reasonable time. In: ISMM.
  pp. 63--74 (Jun 2013)

\bibitem{kaplan98idls}
Kaplan, A., Ridgway, J., Wileden, J.C.: Why {IDLs} are not ideal. In: 9th
  International Workshop on Software Specification and Design (Apr 1998)

\bibitem{kelsey94tractable}
Kelsey, R.A., Rees, J.A.: A tractable {Scheme} implementation. Lisp Symb.
  Comput.  7(4),  315--335 (Dec 1994)

\bibitem{lapaime86logicon}
Lapaime, G., Chapleau, S.: Logicon: An integration of prolog into icon.
  Software---Practice \& Experience  16(10),  925--944 (oct 1986)

\bibitem{mitchell70design}
Mitchell, J.G.: The design and construction of flexible and efficient
  interactive programming systems. Ph.D. thesis, Carnegie Mellon University
  (Jun 1970)

\bibitem{neumann28theorie}
von Neumann, J.: Zur {Theorie} der {Gesellschaftsspiele}. Mathematische Annalen
   100(1),  295--320 (1928)

\bibitem{neumann44theory}
von Neumann, J.: Theory of games and economic behavior. Princeton University
  Press (1944)

\bibitem{rigo06pypy}
Rigo, A., Pedroni, S.: {PyPy's} approach to virtual machine construction. In:
  DLS. pp. 944--953 (Oct 2006)

\bibitem{robinson80loglisp}
Robinson, J.A., Silbert, E.: {LOGLISP}: An alternative to {PROLOG}. Tech. rep.,
  Syracuse University (1980)

\bibitem{sarimbekov13characteriustics}
Sarimbekov, A., Podzimek, A., Bulej, L., Zheng, Y., Ricci, N., Binder, W.:
  Characteristics of dynamic {JVM} languages. In: VMIL. pp. 11--20 (Oct 2013)

\bibitem{sullivan03dynamic}
Sullivan, G.T., Bruening, D.L., Baron, I., Garnett, T., Amarasinghe, S.:
  Dynamic native optimization of interpreters. In: IVME. pp. 50--57 (Jun 2003)

\bibitem{tratt08domainspecific}
Tratt, L.: Domain specific language implementation via compile-time
  meta-programming. TOPLAS  30(6),  1--40 (Oct 2008)

\bibitem{vitek11repeatability}
Vitek, J., Kalibera, T.: Repeatability, reproducibility, and rigor in systems
  research. In: EMSOFT. pp. 33--38 (Oct 2011)

\bibitem{wielemaker12future}
Wielemaker, J.: {SWI-Prolog}: history and focus for the future. ALP  152 (June
  2012)

\bibitem{yermolovich09optimization}
Yermolovich, A., Wimmer, C., Franz, M.: Optimization of dynamic languages using
  hierarchical layering of virtual machines. In: DLS. pp. 79--88 (Oct 2009)

\end{thebibliography}

\onecolumn
\newpage
\appendix

\section{A Larger Example}

\noindent To give readers an idea of what a composed Python / Prolog program looks like,
in this appendix we show an elided example of a program which uses a SAT solver.
It exhaustively enumerates the satisfying models of a SAT instance, counting the number of propositional variables
set true.

The SAT instance is read from a DIMACS file. The solver is called from
Python by invoking the Prolog \texttt{sat} predicate. This converts from a ground
representation where SAT variables are atoms, to one where they are represented
as Prolog variables. It then invokes the Howe/King SAT solver~\cite{howe10pearl}.

\begin{lstlisting}[basicstyle={\footnotesize\tt}]
import uni

class SatSolver(object):
    def __init__(self):
        self.engine = uni.Engine("""
extract_variables([], []).
extract_variables([_/Var | T], [Var | T2]) :-
    extract_variables(T, T2).

variablify(ReifiedProblem, ProblemWithVariables, Mapping) :-
    % convert representation into what the sat solver expects

sat(ReifiedProblem, Result) :-
    variablify(ReifiedProblem, ProblemWithVariables, Result),
    extract_variables(Result, Variables),
    sat_from_paper(ProblemWithVariables, Variables).

sat_from_paper(Clauses, Vars) :-
    % as in the Howe/King paper""")

    # Generate prolog terms to pass to the SAT solver
    def _generate_input_terms(self, py_clauses):
        minus_term = getattr(self.engine.terms, "-")
        pl_cnf = []
        for py_clause in py_clauses:
            pl_clause = []
            for py_lit in py_clause:
                var = "v" + str(abs(py_lit))
                if py_lit < 0:
                    polarity = "false"
                else:
                    polarity = "true"
                pl_lit = minus_term(polarity, var)
                pl_clause.append(pl_lit)
            pl_cnf.append(pl_clause)
        return pl_cnf

    def enumerate_in_python(self, clauses):
        pl_clauses = self._generate_input_terms(clauses)
        counts = []
        for (model, ) in self.engine.db.sat.iter(pl_clauses, None):
            count = 0
            for varname, binding in model:
                count += int(binding == "true")
            counts.append(count)
        return counts

print SatSolver().enumerate_in_python(parse_dimacs(fh))
\end{lstlisting}

\end{document}